# WyCryst: Wyckoff Inorganic Crystal Generator Framework

**Ruiming Zhu[1,2], Wei Nong[1], Shuya Yamazaki[1,2], Kedar Hippalgaonkar[1,2*]**

[1] School of Materials Science and Engineering, Nanyang Technological University, Singapore 639798, Singapore

[2] Institute of Materials Research and Engineering, Agency for Science, Technology and Research (A*STAR), Singapore 138634, Singapore

Correspondence to: kedar@ntu.edu.sg

# ABSTRACT

Generative design marks a significant data-driven advancement in the exploration of novel inorganic materials, which entails learning the symmetry equivalent to the crystal structure prediction (CSP) task and subsequent learning of their target properties. Generative models have been developed in the last few years that use custom Variational Autoencoders (VAEs), Generative Adversarial Networks (GANs), and diffusion models. While periodicity and global Euclidian symmetry in three dimensions through translations, rotations and reflections have recently been accounted for, symmetry constraints within allowed space groups have not. This is especially important because the final step involves energy relaxation on the generated crystal structures to find the relaxed crystal structure, typically using Density Functional Theory (DFT). To address this explicitly, we introduce a generative design framework (WyCryst), composed of three pivotal components: 1) a Wyckoff position based inorganic crystal representation, 2) a property-directed VAE model and 3) an automated DFT workflow for structure refinement. By implementing loss functions that punish non-realistic crystal structures, our model selectively generates materials that follow the ground truth of unit cell space group symmetry by encoding the Wyckoff representation for each space group. In leave-one-out validation experiments, we successfully reproduce a variety of existing materials: $CaTiO_3$ (space group, SG No. 62 and 221), $CsPbI_3$ (SG No. 221), $BaTiO_3$ (SG No. 160), and $CuInS_2$ (SG No.122) for both ground state as well as polymorphic crystal structure predictions for desired compositions. We also generate several new ternary materials not found in the inorganic materials database (Materials Project), which are proved to be stable, retaining their symmetry, and we also check their phonon stability, using our automated DFT workflow highlighting the validity of our approach. We believe our symmetry-aware WyCryst takes a vital step towards AI-driven inorganic materials discovery.

# Introduction

In today's rapidly advancing technological landscape, the need for novel inorganic materials with tailored properties is more pressing than ever. From energy storage and conversion to advanced electronics and catalysis, many modern technologies are driven by materials' innovations. Governed by their unique crystal structures, inorganic materials play a pivotal role in these applications. The quest for predicting stable crystal structures of inorganic materials is, therefore, at the forefront of modern materials science. Quantum mechanical calculations, particularly using Density Functional Theory (DFT), have become the popular standard for predicting and validating the electronic, mechanical, and thermodynamic properties of inorganic materials.[1] DFT provides a balance between computational efficiency and accuracy, making it possible to study materials at the atomic and electronic levels.[2] However, since DFT's structural relaxation and properties prediction relies on known structures, the challenge of using DFT lies in generating new inorganic crystalline materials with unknown crystal structures. Traditional methods, such as evolutionary algorithms, now improved with more efficient optimization strategies, have recently been successful in unveiling new inorganic materials but often come with large computational overheads.[3]

Given the vast compositional space and structure of inorganic materials, a CSP task, especially for finding new materials, can be computationally intensive.[1,4] Generative models, with their ability to rapidly generate potential inorganic structures, offer a more efficient alternative. Central to their success is the concept of an invertible material representation.[1] This bi-directional transformation allows real-world inorganic crystal structures to be converted into a set of features and vice versa. In some generative models, crystal representations of a large dataset of materials are used to construct a lower-dimensional, condensed latent space.[5] Such a strategy ensures that the structures generated by navigating the latent space adhere to the physical and chemical rules governing inorganic materials, making them not just random arrangements of atoms but physically meaningful entities.

Such generative models have produced varying success in computer-assisted drug discovery, particularly for the de novo design of molecules with desired properties.[6-9] These models can automatically generate new, previously unexplored bioactive and synthesizable molecules in a time- and cost-effective manner. Some examples of success stories of generative models in small molecules for energy and drug discovery include the identification of lead compounds for drug development and the discovery of new materials for energy storage.[8,9] However, there are still challenges that need to be addressed, such as the need for more diverse training data and the development of more efficient algorithms for generating molecules with specific properties.[8] The generative design of inorganic materials similarly involves a key step of identifying invertible representations of structural features that are trained on sufficient, curated and representative materials data coupled with carefully curated models that capture the scientific knowledge required for the construction of crystalline materials.[1,4] The success of such a paradigm would effectively replace or augment the standard, yet expensive DFT calculations for CSP tasks.

Given materials data with properties of interest, the first step to generative design is to establish the invertible feature representation. For small organic molecules typically consisting of few main group elements from the periodic table, string-based representations such as SMILES[10] and SELFIES[11] along with graph representations[12-14] have been established with language or transformer-based machine-learning (ML) models applied. Inorganic crystalline materials are more diverse in chemical composition, but they have limited structural variations, resulting in fewer available robust representations. Typically, VAEs, GANs, diffusion models, recurrent neural networks (RNNs), and other hybrid models, when coupled with various feature representations, have demonstrated generative capabilities. The VAE-based iMatGen is reportedly the first model to achieve inverse crystal structure generation of vanadium oxides by encoding the crystal unit cells and atomic positions into an image representation of three-dimensional (3D) grids. However, it is limited to the V-O binary system, and its performance is computationally constrained by memory-intensive training from 3D voxel image representations.[15] Another model based on voxel crystal representation is the conditional deep-feature-consistent VAE, which predicts eight target properties for various typical cubic compounds.[16] The Fourier-transformed crystal properties (FTCP) representation for VAE model is a more generalized approach for broad crystal systems, capable of inverse properties-structure design by adding a target-learning branch, and efficient structure generation and high match rate albeit with relatively lower structure validity [17]. Recently, diffusion models have emerged as a potential solution, e.g, crystal diffusion variational autoencoder (CDVAE)[18] and its successor MatterGen[19], offering an E(3) equivariant modeling approach to generating structures by iteratively refining random initial configurations until they match the desired distribution. The combination of diffusion models and graph neural networks enables both CDVAE and MatterGen to maintain SE(3) invariance. DiffCSP achieves both E(3) and O(3) invariance by incorporating constraints during the diffusion process. The enhanced version, DiffCSP++, not only refines the symmetry constraints in the diffusion process for E(3) and O(3) invariance, but also introduces constraints on the Wyckoff positions of the fractional coordinates.
Ensuring a balance among various models and techniques is vital to generate valid and diverse inorganic structures, pushing the boundaries of what we know about inorganic materials.

GAN models have been used with varying success. CrystalGAN was the first attempt using a GAN to generate ternary hydrides from binary observations while no evidence is available if it can be generalized to other broad crystal structures[20] The composition-conditioned crystal GAN using point cloud representation generates structures of a fixed ternary compound space (Mg−Mn−O) with the ability to predict new compositions with band gaps relevant for photocatalysis.[21] A constrained crystals deep convolutional GAN (CCDCGAN) using voxel representation for lattice constants and atomic positions and two-dimensional (2D) crystal graph autoencoder, was developed to explore Bi-Se binary chemical space.[22] MatGAN employed a matrix representation that one-hot encodes compositions of crystals and generates a large amount of new hypothetical materials[23], while the validation from the perspective of DFT stability or experimentally was missing. Improving upon this, CubicGAN was shown to successfully generate diverse cubic multi-component materials and four new prototype cubic materials were further filtered down, with which close to hundred stable materials were confirmed stable using phonon dispersions over three DFT screening steps.[24] However, this model was constrained to cubic symmetry for both trained and generated crystals into only three space groups, i.e., No. 216, 221 and 225 out of 36 possible cubic space groups (No. 195 to 230) even within cubic symmetry. This limits the structural diversity of the model.

Recently, hybrid models combining machine learning with DFT relaxation also have been developed. For example, a deep learning-based physics guided crystal generative model (PGCGM) includes the losses of atomic pairwise distance and structural symmetry, leading to increased validity in structure generation for inorganic ternary materials covering 20 individual space groups, with results comparable to FTCP: however, the optimization step significantly changes the symmetry upon energy relaxation.[25] Along these lines, a RNN generative model using an invertible string representation has recently been developed, where a simplified line-input crystal-encoding system (SLICES), was developed showing invertibility with a high reconstruction match rate for 40,000 crystal structures that hold symmetry invariances.[26] The symmetry invariance is realized over two consecutive steps of structure refinement via machined-learned force-field and inter-atomic potentials. CrysTens is an image-like representation that encodes the lattice, basis and atomic positions, matrix-form pairwise distances[27], was claimed to be able to combine with GANs and diffusion models for crystal structure generation, identifying 6 new crystals out of top 35 structures with predicted low formation energy from 1, 000 generated outputs, during which the manual tuning (rounding) of atomic positions and atomic numbers was adopted. Trained on huge number of text-form crystallographic information files, a transformer-based large language model named CrystaLLM shows capability of new crystal generation[28]. DiffCSP[29] and DiffCSP++[30], which are models grounded in the diffusion model and space group symmetry, produce structures that comply with symmetry by imposing explicit symmetry constraints in the diffusion process. All the above approaches, while promising, have their drawbacks: VAEs might produce inorganic structures with questionable validity, while GANs can be tricky to train, especially when dealing with the complex nature of inorganic materials. Issues like non-convergence and mode collapse can arise. Diffusion models still suffer from inefficient sampling speed for structure generation compared to VAE models. The CrystaLLM, due to its large language modeling (LLM) nature, requires a substantial amount of training data and computational resources, making it a costly model to train and fine-tune. DiffCSP++ necessitates the use of Density Functional Theory (DFT) calculations to verify the stability of the materials it generates.

Most importantly, a common, critical issue exists in most structure generation attempts: the sampled (generated) crystal structures need to conform to crystallography principles that fundamentally requires the output structures to be physically meaningful. The most effective approach thus far involves incorporating Wyckoff site-symmetry constraints into the generation process using diffusion models, as demonstrated in previously mentioned DiffCSP++. Other models produce "random" structures without 100% validity and most generated structures undergo non-trivial DFT relaxations ("cannot be optimized"). Additionally, the symmetry of the remaining reasonable structures could be altered during DFT relaxation, i.e., change of generated symmetry characteristics, which conflicts with desired property-targeted inverse design as property emerges directly from generated structure.

Our WyCryst framework offers a robust solution to the challenges in generative design of inorganic materials. It employs a Wyckoff Position-based representation that explicitly accounts Wyckoff site symmetry for each space group, ensuring the symmetry is retained not only in the generation process but also after subsequent structure optimization. Furthermore, the Property-directed VAE model with design to penalize non-adherence to symmetries within the space groups ensures that the WyCryst framework generates crystals respecting these symmetry rules. Complemented by an automated DFT workflow, WyCryst refines these structures, therefore maintaining their desired properties. This approach not only reproduces known materials but also introduces novel, stable ternary compounds, effectively bridging generative design with practical applications. Notably, in materials science, structure defines property. Hence, by explicitly generating symmetry-aware structures, we are closer to achieving AI-driven materials discovery.

# Workflow

In this section, we present our WyCryst Framework. As shown in Figure 1, WyCryst takes materials Crystallographic Information File (CIF) input and uses an open-source python library for crystalline material structures and analysis, namely Pymatgen [31] to analyze the crystal Space Group (SG) symmetry and Wyckoff site occupancy of all atoms in each unit cell. For any given material in the training set encompassing $N$ materials, the above information is processed into a Wyckoff representation $\{W_i = (X_i, S_i)\}_{i\in[N]}$, where $X_i$ and $S_i$ are the corresponding sets of Wyckoff arrays and one-hot encoded SG arrays. We make use of the Property-directed Variational Autoencoder (PVAE) to learn the overall distribution of inorganic crystal materials in our probability distribution. The latent space in the PVAE is also connected to one or several target properties through a fully connected neural network, which guides the latent space to have a gradient with respect to physical properties of materials. For materials reconstruction/generation, a decoder is employed to map the latent space arrays back to original sets of Wyckoff representation arrays. These represents the irreducible building blocks to generate the lattice and basis vectors for inorganic crystals in a dataframe format that we henceforth call the 'Wyckoff Genes', which include 3 key components: composition, space group and corresponding Wyckoff site occupancy. Ultimately, we design an automatic CIF generation and DFT optimization workflow to generate the final output crystal materials. In the following subsections, we introduce each component of WyCryst and how this workflow enables the learning of crystal symmetry information and new material generation.

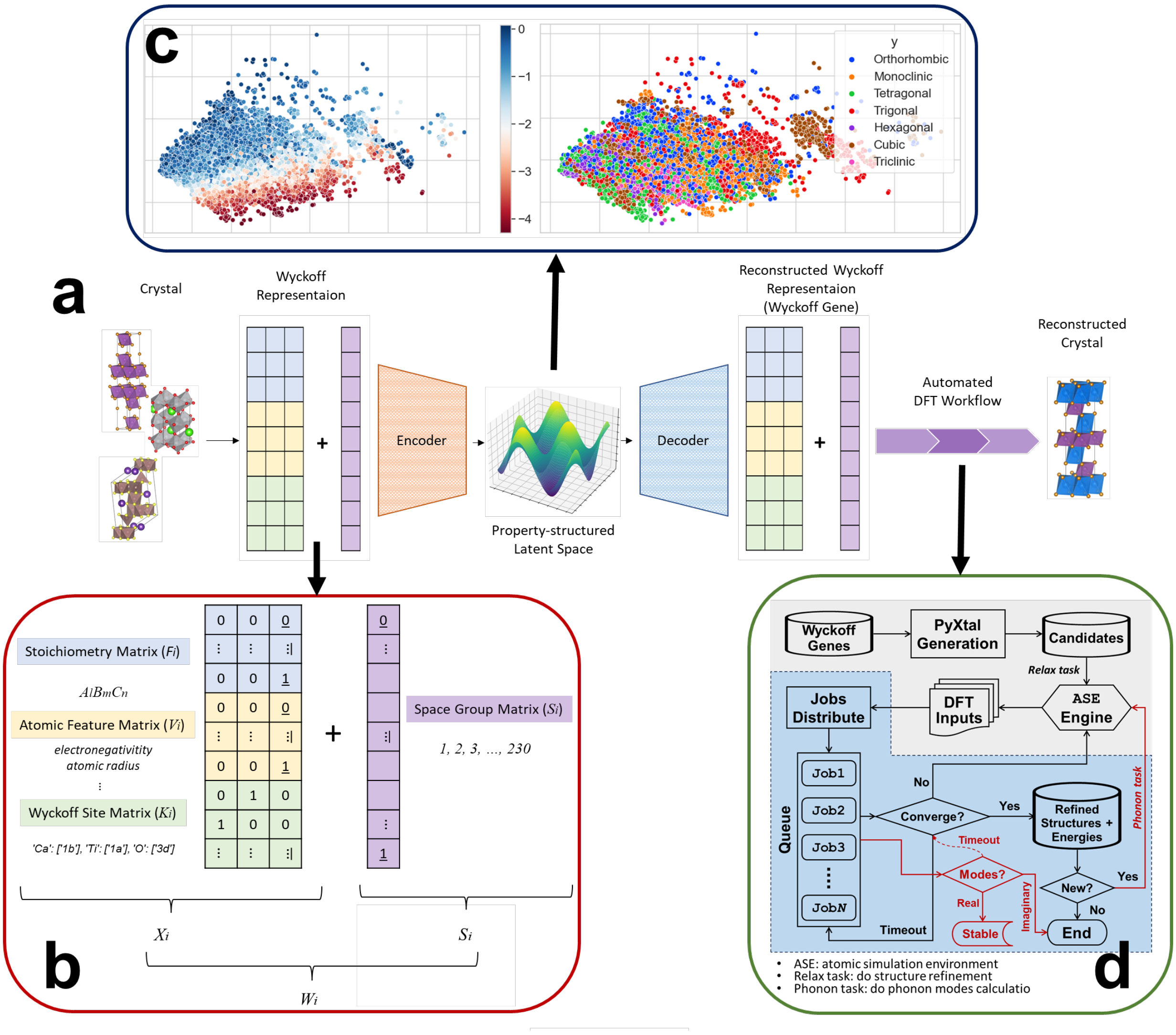

Figure 1: (a) Wyckoff Inorganic Crystal Generator Framework including: (b) Wyckoff-based Feature Representation composed of stoichiometry, atomic features, crystal space group and the corresponding Wyckoff positions; (c) Principal Component Analysis (PCA) visualization of latent space of the PVAE model with two different labels: formation energy and crystal systems; (d) DFT workflow including crystal structure generation, energy relaxation, refinement, and stability check.

## Crystal symmetry, Wyckoff, and equivalent Wyckoff positions

Inorganic crystal materials can be characterized by their distinct chemical formulas (compositions) and precisely arranged periodic structures (unit cells). A repeating periodic unit (called a lattice), when combined with the position of atoms within the unit cell (basis) constitutes the material. Within each unit cell, well-defined lattice parameters denoted by 'a', 'b', and 'c' axes, along with the corresponding angles ‘α’, 'β', and 'γ', are established. These parameters must adhere to the principles of symmetry, playing a fundamental role in shaping the properties and behaviors of these materials. The cornerstone of symmetrical determination in such materials lies in their designated space group, which serves as the primary framework for describing their symmetrical arrangements. In each space group, symmetry operations can be further dissected into distinct Wyckoff sites[32,33], which the atoms occupy. Each Wyckoff site of each space group occupies fixed number of specific positions of the unit cell according to the predefined symmetry operations. Furthermore, most space groups have one or several different but equivalent sets of Wyckoff positions, which are interchangeable with certain rotation and translation symmetry operations. Equivalent Wyckoff positions are important for understanding crystal symmetry, which we will address in our validation section. In addition, within our crystal generation and Wyckoff Gene constructions, we introduce a key parameter: degrees of freedom (DoF). For each space group, since the three-dimensional symmetry operations (such as translation, inversion and rotation) are known, a single general position exists with DoF = 3, in addition to a finite number of special positions (DoF = 0, 1, or 2) that are progressively invariant to the allowed symmetry operations. In our Wyckoff Gene, DoF symbolizes the total number of unknown parameters to optimize during the relaxation process, especially in the DFT calculations for refinement. We use DoF as an approximation of the complexity of the ground state crystal structure prediction task.

**Wyckoff Position-based Inorganic Crystal Representation**

We develop a crystal representation based on the fundamental crystal symmetry rules, which contains two main components: space group number $S_i$ and Wyckoff array $X_i$ as shown in Figure 1(b). The space group array $S_i$ is a one-hot encoded label matrix with dimension of total number of space groups $\{230\}$. Wyckoff array $X_i = (F_i, V_i, W_i)$ contains stoichiometry and Wyckoff symmetry site information of the material. $F_i$ denotes a one-hot encoded element stoichiometry matrix that symbolize the crystal material chemical formula $\{A_l B_m C_n\}$. $V_i$ denotes the atomic features matrix proposed in the crystal graph convolutional neural network (CGCNN)[34]. Finally, $W_i$ denotes the Wyckoff site occupancy and Wyckoff site multiplicity of each element in the material. The combination of $S_i$ and $W_i$ describe Wyckoff position constraints of the fractional coordinates. In this representation, we focus on the invertibility of the representation, which necessitates minimal energy and structure change to go from generated representation to the final structure (fully DFT relaxed). The one-hot encoded stoichiometry matrix $F_i$ allows us to reconstruct the chemical formula while the space group array $S_i$ and Wyckoff $W_i$ ensures the generated crystals comply to symmetry rules.

**Property-directed VAE Model**

The PVAE features two main objectives to enable the construction of new materials: 1) learning the distribution of inorganic crystal materials from the database, and 2) creating a property-oriented latent distribution through a property learning branch. These objectives are realized by firstly the encoder of PVAE, which uses a convolutional neural network (CNN) to map input Wyckoff arrays into a lower-dimensional latent space. Specifically, the encoder outputs $Z_{mean}$ and $Z_{variance}$, which parameterize a multivariate Gaussian distribution in the latent space. Secondly, the decoder is trained to generate samples within a range of variability around the latent point's mean ($Z_{mean}$) and to reconstruct this neighborhood of $Z_{mean}$ back into the original Wyckoff representation, generating new materials with their representative Wyckoff Genes. During this process, two loss functions are implemented: reconstruction loss $L_{recon}$ and KL (Kullback-Leibler) divergence loss $L_{KL}$. The reconstruction loss encourages the model to generate output data that is as close as possible to the original input data. The reconstruction loss is a combination of the mean squared error (MSE) and the cross-entropy loss:

$$L_{recon} = MSE(X, X') + H(S, S') \quad (1)$$

where $(X, X')$ *and* $(S, S')$ are the input and reconstructed Wyckoff array and space group array. It is important to note that the atomic feature matrix is included when calculating the reconstruction error, even though these features may not always correspond with the stoichiometry matrix upon reconstruction part and do not get processed toward subsequent Wyckoff Gene generation. These elemental features play a crucial role in providing the model with additional information about the material during forward prediction and reconstruction tasks. This characteristic is not unique to our model but is also observed in other generative and forward models, as it is a fundamental aspect of the design of Variational Autoencoders (VAEs) and other generative models that incorporate atomic features as part of the input.

The KL loss in PVAE is used to shape the latent space into a Gaussian distribution which helps regularize the latent space and ensures a continuous distribution. Mathematically, if $q(z|x)$ represents the learned distribution of latent points given the input data $X$ and $p(z)$ is the desired Gaussian distribution, the KL loss is calculated as:

$$L_{KL} = KL(q(z|x)||p(z)) \tag{2}$$

We also implemented a property-learning branch which connects the latent space to one or several target properties. Through learning of specific material properties, such as formation energy, bandgap energy, bulk modulus, and others, PVAE enables property-sloped latent space. The learning is achieved using the property loss:

$$L_{prop} = MSE(Y, Y_{predicted}) \tag{3}$$

where $MSE(Y, Y_{predicted})$ is the sum of MSE between true values and predicted values for all targeted properties. Lastly, we present one final, and critical, loss function to ensure symmetries within the space group are obeyed for each generated material. This Wyckoff loss $L_{Wyckoff}$ calculates the mean square error between the original formula (or generated formula, depending on the setting), and the reconstructed symmetry-constrained Wyckoff-weighted formula:

$$L_{Wyckoff} = MSE(R_{True}, R_{Wyckoff}) \tag{4}$$

where $R_{True}$ denotes the input formula and $R_{Wyckoff}$ denotes the reconstructed formula. In this computation, both the integrity of the initial space group symmetry details is preserved, and efforts are made to reduce discrepancies on Wyckoff sites without physical meaning. Notably, our symmetry-compliant WyCryst approach is better than previous models in that atomic positions and angles are not necessary in our generative PVAE model for producing stable crystal materials; instead this prediction is implicit due to the matching of symmetry considerations. Essentially, the advantage is two-fold: (1) The angles can be 'snapped-to-grid' based on this learnt crystal symmetry and (2) the exact lattice constants (given all possible permutations that obey the symmetry rules of the generated space group) will be learnt through a computationally inexpensive DFT refinement step described below.

**Automated DFT Workflow for Structure Refinement**

The Wyckoff Genes generated by our PVAE model are entered into the DFT workflow which generates possible crystal structures. The first step is to ensure that only realistic and practical structures are inputted into DFT calculations. There are two primary categories of filters utilized: filters for reliability and filters for experimental feasibility. The reliability filters consist of filters for chemical formulas, checkers for Wyckoff positions, and filters for duplicate Wyckoff Genes. On the other hand, the experimental feasibility filters include filters for elements, preferences for "space group + DoF", and filters for synthesizability. Building on these symmetry-compliant and experiment feasible Wyckoff genes, some guesses of atomic positions are needed for DFT calculations to perform structure refinement to ascertain final lattice parameters. We employed PyXtal[35], a Python library for possible crystal structure generation, to produce the initial unrelaxed crystal structures, in which the lattice parameters and atomic positions of the unrelaxed structures satisfy the lower limit of the interatomic distance estimated by the corresponding atomic covalent radii. For Wyckoff Genes with non-zero total DoF, we set the number of trial structures as (total DoF + 1) per Wyckoff Gene, which from our tests is sufficient to capture the variation caused by the DoF of each Wyckoff position. For the case of DoF = 0, a unique collection of atomic positions can always be generated as all atoms are located in exact positions, a key observation for higher symmetry structures. All the generated symmetric crystal structures (candidates) are then relaxed using Vienna ab-initio simulation package (VASP)[36] through the input files preparation by the atomic simulation environment (ASE) engine[37]. Within the refinement process, lattice parameters and atomic positions are allowed to be relaxed, based on their DoFs but with the constraint of space group symmetry, using the standard DFT calculations. More details on DFT calculations are given in the Supplementary Information. Those relaxed structures accompanied with energies were filed into disk and analyzed for novelty, if they are different from the training materials database (MP)[38]. The new materials are subsequently checked for dynamic (phonon) stability through the implementation of density functional perturbation theory (DFPT) calculations: here, a stable material requires that all phonon modes do not have any imaginary component[24]. The full framework is shown below in Figure 1(d). Note that we also tried structure refinement using the machine-learned interatomic potential, e.g., M3GNET[39], a proposed universal interatomic potential that are trained on the data of DFT relaxation trajectories, with lower computational demands. However, as shown in Table S1, the validation test on $CaTiO_3$ structure refinement shows that using M3GNET leads to a symmetry change from space group No. 62 to No. 14 without the symmetry constraint or the force divergence with the symmetry constraint, while the original symmetry of crystal structure is preserved and the forces converge when we use traditional DFT. Structure determines property and hence, such symmetry invariance in the refinement process is critical for the generative design of materials with target properties: this requirement justifies the use of relatively high-fidelity DFT calculations for the structure refinement.

# Results and Discussion

In this section, we evaluate the potential of WyCryst in generating new materials using the open-source Materials Project (MP) Database. We first test the performance of Wyckoff representation based forward model on targeted material properties. Secondly, we show the reconstruction quality and property prediction accuracy of the PVAE model. We also plot the property-structured latent space to prove the effectiveness of the property learning branch and the overall model structure. Next, we perform several leave-one-out validations to demonstrate how the WyCryst workflow successfully reproduces existing materials. Lastly, we sample new ternary materials using a custom element list, which are earth-abundant materials easily accessible to experimental labs. These generated materials are not in the MP database and are proved to be stable using our automated DFT workflow.

The datasets for all experiments are obtained from the MP. We queried a total of 66,643 ternary compounds from the MP database (the database version we used is v2023.7.4). The distribution of space group numbers and atomic numbers of our training set are plotted in the supplementary information. The training set includes materials from all seven crystal systems and all elements up to atomic number 83 (Bismuth), which are sufficient for our model to learn the latent distribution and to sample new materials (Supplementary Information section 8). We run all the model training and data sampling on a laptop platform with the following configuration: Intel Core I9-13980HX, 64GB DDR5 4800MHz RAM, and NVIDIA GeForce RTX4090 Laptop GPU with 16GB VRAM. We run DFT calculations on a server platform with the following configuration: Intel Xeon Phi 7210 64-cores @1.3 GHz, 192GB DDR4 2133MHz RAM. The computation times of each step are summarized in Table S2. The deep learning model training, latent space sampling, and generation of crystal structures are computationally low-cost. While the subsequent structure relaxation and phonon stability check consume significantly more computation resources, this is common for the implementation of DFT calculations and is relatively much faster than a global search for minimal energy structure since in our case, we are only looking for small relaxations with fixed symmetry. Hyperparameters of all models in the experiment section are tuned using Grid Search Cross-Validation, and the final hyperparameters are shown in our code.

**Forward Model Performance**

We performed all the forward regression model experiments using three different crystal representations on three different properties that can be queried from the MP database. Formation energy ($E_f$) is one of the primary properties for considering stability of a crystal, while bandgap energy ($E_g$) is the most important property for semiconductors. Thirdly, going beyond stability, in order to determine if a material can be experimentally synthesized in the laboratory or not, we introduce our recently published synthesizability score (SC) metric.[40] Since FTCP and CGCNN models, as described earlier, serve as a good benchmark, we then compared our forward model based on the Wyckoff representation with them. We used 5-fold cross-validation to test the performance of all three forward models. Table 1 shows the mean absolute error (MAE) and recall performance of three materials properties. Our Wyckoff representation model outperforms CGCNN in predicting formation energy, bandgap energy, and achieves significantly better performance in predicting material synthesizability score (SC). Additionally, our model demonstrates comparable performance to the FTCP's target learning branch, which requires more complex and comprehensive analysis of the crystal, as it not only incorporates real-space features but also includes reciprocal space characteristics. It's noteworthy that recent graph-based or attention models such as MEGNET[41], ALIGNN[42], CRABNET[43], and others have achieved formation energy prediction performance in the range of 0.026-0.039eV/atom. However, the performance of the forward model is not as crucial as the performance of the subsequent generative model. The main reason for presenting these results is to show that our model is comparable to recent advancements in forward model work and that the Wyckoff representation is well-suited for the subsequent workflow, and the forward model performance can be fine-tuned further if necessary. Leveraging the Wyckoff representation, which summarizes crystal information by emphasizing crystal symmetry rather than exact atomic positions, yields comparable or potentially enhanced performance levels. Thus, our performance demonstrates that the critical information about the material is embedded in the symmetries for the crystal structure, which are sufficient to describe the stability, band structure and synthesizability of a material. And indeed, our Wyckoff representation is an excellent choice for both forward learning models and invertible generative models.

| Property | CGCNN | FTCP | WyCryst |
|---|---|---|---|
| $E_f$ (eV/atom) | 0.055 | 0.051 | 0.044 |
| $E_g$ (eV) | 0.250 | 0.204 | 0.202 |
| SC (recall) | 75.3% | 85.2% | 90.7% |

Table 1: Comparison of two benchmark forward models' performance with WyCryst in predicting three major inorganic crystalline material properties.

### Property-Directed VAE (PVAE) Generative Model Performance

Beyond the forward model accuracy, in order for the generated crystals to be realistic and realizable (especially as an input to the DFT refinement workflow as described in Figure 1(d) above) we have to ensure good reconstruction quality of our model. Again, we benchmarked our WyCryst PVAE with the reconstruction performance of two other state-of-the-art generative VAE models. All VAE experiments are based on the same dataset used in the previous sections, with 5-fold cross-validation. Real-space features VAE is a VAE model based on lattice-only features from the FTCP model, without considering the Fourier transformed features. Table 2 shows the reconstruction accuracy/error of different features in all three models. Our PVAE is more accurate in reconstructing the elements making up a ternary formula and exhibits high Wyckoff site and Space Group accuracies (both higher than 90%). Note that our PVAE does not explicitly include the lattice parameters in the loss function as it will be addressed in the following automated DFT workflow section. We also evaluated our model based on the previously defined validity, which is defined as the percentage of generated materials that satisfy the following criteria: (1) the percentage of materials that pass through DFT relaxation (proposed by Ren et al.[17]) and (2) the shortest distance between any two atoms > 0.5 Å (proposed by Xie et al.[18]). FTCP achieved 92.2% validity rate under the first defined metric, while generative models have validity performance ranging from 70%-100% under the second metric. On contrary, Our WyCryst framework achieved 100% validity under both metrics because of the nature of our model requiring symmetry-consistency throughout the entire workflow. However, we'd like to note that these previously defined validity metrics are far from sufficient conditions for a stable and/or synthesizable crystalline material, which we will address in more detail in the material generation section below. We compare the state-of-the-art generative models with all proposed metrics (Supplementary Information section 10).

Next, the Principal Component Analysis (PCA) visualization of the 256x1 latent space (the size is learnt through hyperparameter tuning – see Supplementary Information for details) is shown in Figure 1(c) above. As the property-learning branch in our WyCryst framework uses the first property, namely the formation energy ($E_f$), the latent space visualization shows a gradient with respect to $E_f$. In Figure 1(c), we also plot the distribution of the chemical space by means of showing the 7 crystal systems, which encompass the 230 Space Groups for all inorganic materials. Despite the fact that the latent space is not constructed based on the crystal symmetries and the PCA- analyze latent space can also be significantly affected by the chemical formula of the crystals, we still see clustering of materials with the same crystal systems especially on the top right corner where some cubic and trigonal materials are clustered together. Beyond our current implementation, one could also explore multiple property-learning branches simultaneously, although this will affect the specific reconstruction losses in non-obvious ways and we reserve this for future work. With this high performance in both the forward model and the reconstruction losses, we now proceed with the validation and generation steps.

| Reconstruction Performance | Real-space Features VAE | FTCP [17] | WyCryst PVAE |
| --- | --- | --- | --- |
| Element (accuracy) | 98.1% | 99.0% | 99.9% |
| Wyckoff Site (accuracy) | / | / | 92.6% |
| SG accuracy | / | / | 91.8% |
| Lattice Constant (MAPE) | 12.5% | 9.01% | / |
| Lattice Angles (MAPE) | 8.12% | 5.07% | / |
| Atom Coordinates (MAPE) | 0.047% | 0.045% | / |
| Validity (Ren et al. defined) [17] | / | 92.9% | 100% |
| Validity (Xie et al. defined) [18] | / | / | 100% |

Table 2: Reconstruction performance of two benchmark VAE models compared with WyCryst.

**Leave-One-Out Validation**

The target of the first validation step is to reconstruct a known material by leaving it out of the model training: this itself is a multi-level, difficult challenge because a known material consists of the stoichiometry, atomic information, correct space group and it's corresponding Wyckoff positions, which together constitute the Wyckoff Gene. We simulate this scenario by doing several leave-one-out validation experiments. In each experiment, one target material with specific structure (for example, $CaTiO_3$ with space group 221 and its related Wyckoff site occupancy) is removed from the training set, as if it doesn't exist. Then this leave-one-out dataset is split into 80:20 train-validation sets which are used to train and validate the WyCryst framework. For each left-out material, a separate PVAE model is trained using the same sets of hyperparameters optimized for the entire dataset using Grid Search Cross-Validation; for the purpose of this work, we tested five such materials ($CaTiO_3$, $BaTiO_3$, $SrTiO_3$, $CsPbI_3$, $CuInS_2$) belonging to different space groups. After all these leave-one-out models are trained, we further divide the experiments into two distinct tasks, which are critical from a materials design perspective: (1) Ground State CSP task and (2) Polymorph CSP task. The ground state CSP task validates the WyCryst's ability to predict the most stable crystal structure (lowest formation energy) given a particular materials stoichiometry. On the other hand, the polymorph CSP task focuses on finding polymorphs of a fixed stoichiometry, especially under a specific space group. We attempt to solve two CSP challenge to prove the robustness of WyCryst framework.

**Ground State CSP Task – given a target stoichiometry, can you predict the ground state structure?**

Two typical oxides $CaTiO_3$ and $BaTiO_3$ are chosen as example test cases. The ground state of $CaTiO_3$ below ~ 1500 K under ambient pressure is a tetragonal perovskite in space group No. 62.[44] Below ~ 100 K under ambient pressure $BaTiO_3$ is rhombohedral in structure with a space group of No. 160.[45]

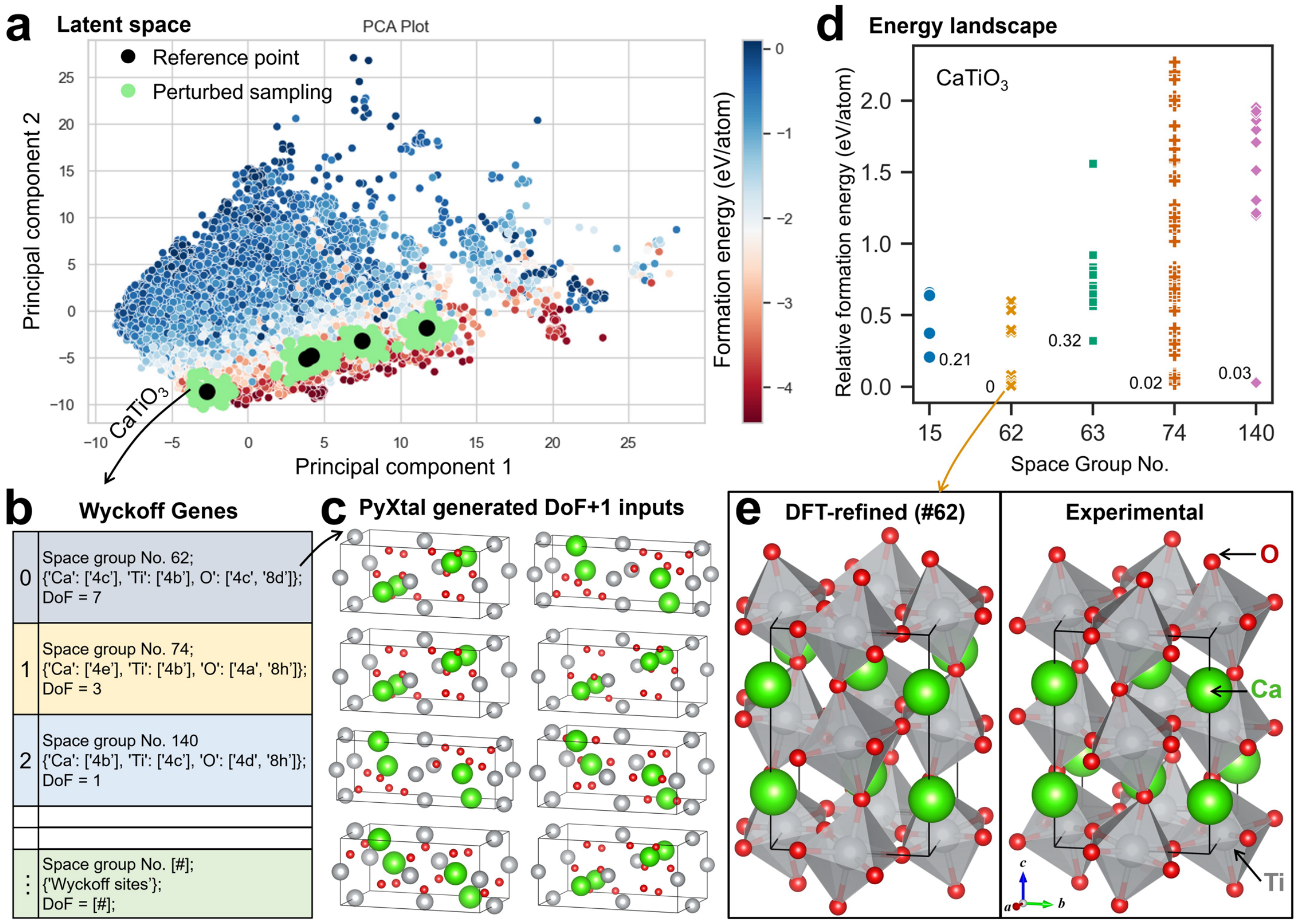

Figure 2: Overview of the ground state CSP task including latent space sampling, Wyckoff Genes, and DFT workflow results: (a) PCA visualization of the $CaTiO_3$ (SG 62) latent space sampling, where black reference points are all materials with the same stoichiometry used as the origin point for sampling and green points are all sampled data points; (b) Wyckoff Genes of sampled materials after post-processing; (c) all PyXtal generated crystal structures from the first Wyckoff Gene (space group 62); (d) The relative formation energy $E_f$ of all 132 refined CaTiO3 structures; (e) crystal structure comparison between WyCryst generated ground state structure and experimental stable structure reported in literature.

To generate Wyckoff Genes for the $CaTiO_3$ CSP task, we sample data points in the property-structure latent space using all ternary compounds with same stoichiometry in the Ca-Ti-O chemical space as origin points (black points in Figure 2(a)). A local perturbation method is used with Gaussian noise whose levels are determined via hyperparameter tuning around the aforementioned origin crystal points. We generated 37 Wyckoff Genes in five space groups (No. 15, 62, 63, 74 and 140) from the leave-one-out PVAE model (details in the Supplementary Information Table S3). Totally 132 trial candidates were produced using PyXtal, from the 37 Wyckoff Genes followed by refinement and DFT relaxation (see Figure 1(d) for workflow). As illustrated in Figure S3, the DFT-refined structures for each Wyckoff Gene in Figure 2(b) are uniquely the same, no matter how different the initial PyXtal generated inputs are (Figure 2c). This proves that the Wyckoff representation is a one-on-one mapping between the property and the crystal structure. The formation energy values, $E_f$, of refined structures were calculated (details for formation energy calculations in the Supplementary Information) and the relative $E_f$, referenced to the lowest one, are shown in Figure 2(d). The lowest $E_f$ is given by the structure in space group No. 62 (real-space representation in Figure 2(e), which represents the ground state crystal structure and matched the experimental crystal structure). The refined lattice parameters of this structure are $a$ = 5.51 Å, $b$ = 7.69 Å, $c$ = 5.37 Å, and $\alpha = \beta = \gamma$ = 90°, which are consistent with the experimental results $a$ = 5.44 Å $b$ = 7.64 Å, $c$ = 5.37 Å, and $\alpha = \beta = \gamma$ = 90°[46]. The structure features such as the bonding and the co-ordination environments of the atoms also confirm that it is the ground-state $CaTiO_3$. Similarly, for the CSP of ground-state $BaTiO_3$ (Figure S1 and Table S4), 126 trial candidates in 11 space groups were generated using 26 Wyckoff Genes. After implementing our automated DFT workflow, $BaTiO_3$ in space group No. 160 was found to have the lowest $E_f$ and its lattice parameters are consistent with the experimental observations ($a = b = c$ = 4.00 Å, $\alpha = \beta = \gamma$ = 89.8°)[47]. The experimental and DFT-refined crystallographic information of $CaTiO_3$ and $BaTiO_3$ are given in Table S5, further confirming the identicality.

**Polymorph CSP Task – does a polymorph of a desired stoichiometry exist with desired space group symmetry?**

Here, we focus on a technologically relevant portion of phase prediction of high-symmetry structures, in a similar leave-one-out approach. Hence, high-temperature simple cubic phases (space group No. 221) of $CaTiO_3$, $SrTiO_3$ and $CsPbI_3$ and a typical thermoelectric material $CuInS_2$ in tetragonal phase (space group No.122) were chosen as the validation targets for this task. The sampling strategy of the latent space is similar to the one described the ground state CSP task, except now we sampled from all existing ternary materials from the database belonging to the target symmetry. Figure 3 shows the latent space sampling for all four leave-one-out materials with their generated Wyckoff Genes and final DFT relaxed crystal structure. We successfully recreated all four materials in their original Wyckoff configurations (with two more equivalent configurations for $CaTiO_3$ and $SrTiO_3$, which will be addressed in the next section). The final refined lattice constants for $CaTiO_3$, $SrTiO_3$ and $CsPbI_3$ are 3.89, 3.94 and 6.36 Å, agreeing with experimental results[44,48,49]. As for $CuInS_2$, $a = b$ = 5.58 Å, $c$ = 11.24 Å, are consistent with experimental observations ($a = b$ = 5.52 Å, $c$ = 11.13 Å)[50]. This shows the strength of our WyCryst framework as the validation of both ground-state and polymorphic crystal structures is successful. All the reconstructed Wyckoff Genes can be found in Supplementary Information Table S6.

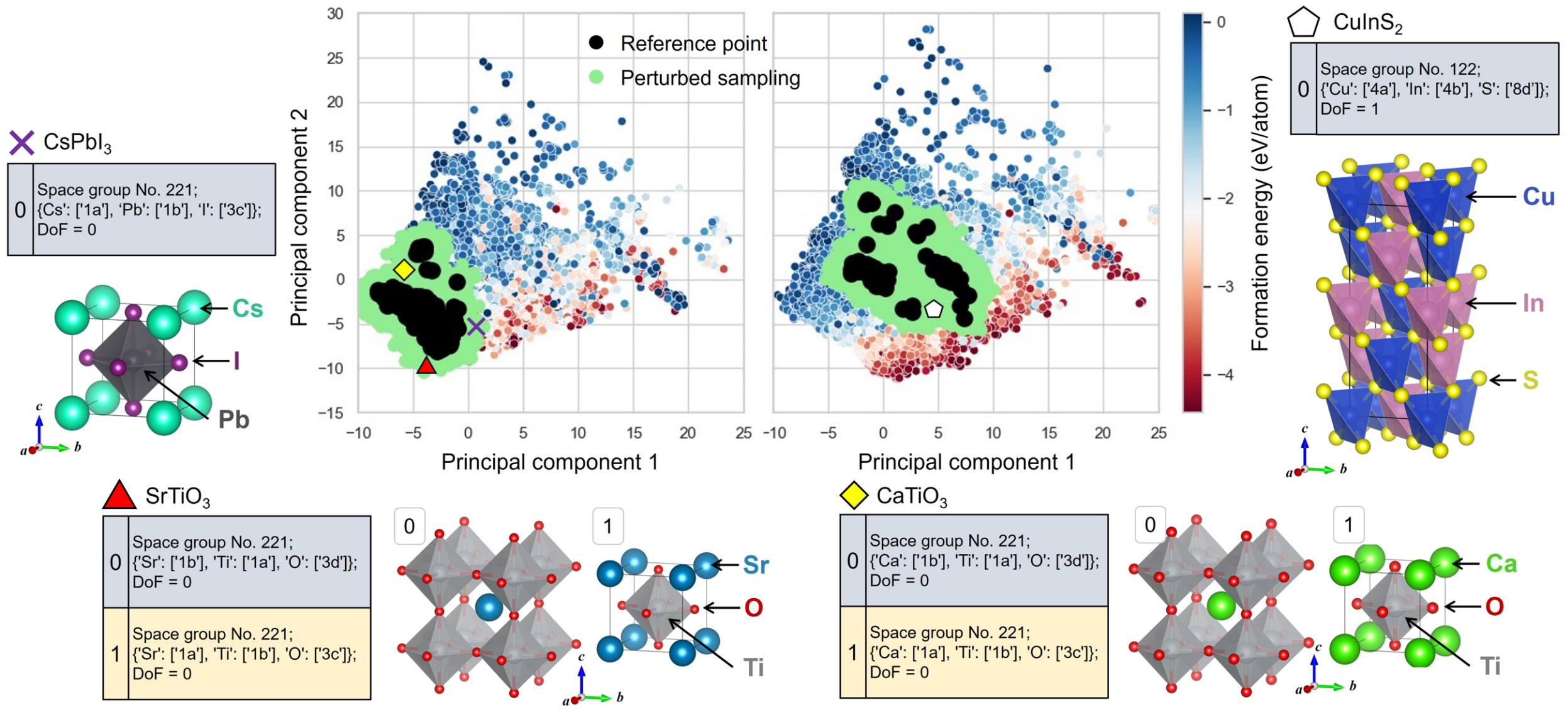

Figure 3: Visualization of the polymorph CSP task including latent space sampling, generated Wyckoff Genes, and unrelaxed crystal structure for all four leave-one-out materials: $CaTiO_3$, SrTiO3, $CsPbI_3$ and $CuInS_2$. Here, we used the PCA plot for $SrTiO_3$ (left) and $CuInS_2$ (right) to illustrate the sampling process and the actual latent space sampling for each material are performed with individual models (shown in Figure S2).

**Equivalent Wyckoff Genes**

In the preceding validation experiments, we have successfully reproduced all leave-one-out materials. Surprisingly, we observe that our WyCryst generated Wyckoff Genes occupy not only their expected Wyckoff position configuration (same as the leave-one-out configuration), but also in their equivalent sets of Wyckoff position. Table 3 shows all Wyckoff Genes that are generated with more than one Wyckoff position configurations. For instance, in validation of CaTiO3 with SG 62, the generated Wyckoff Gene encompasses two distinct sets of Wyckoff position configurations, where Titanium atoms are occupying either '4a' or '4b' Wyckoff positions in two equivalent configurations. These two Wyckoff Genes are interchangeable through a specific set of symmetry operations: a translation operation [0, 0, 1/2] along the Z direction. Importantly, this phenomenon is not unique to $CaTiO_3$ during leave-one-out validations but extends to other materials, including $CaTiO_3$ (SG 221), $SrTiO_3$ (SG221), and $Zn_2GeSe_4$ (SG122). Taking $SrTiO_3$ as another example, Figure 3 shows the pair of equivalent Wyckoff Genes in the validation experiment. Two generated structure unit cell visualizations are different but indeed identical crystals. The capability to identify equivalent Wyckoff positions without explicit training on them serves as compelling evidence that our Wyckoff representation, in conjunction with the PVAE, possesses an understanding of crystal symmetry. Specifically, the sampling from latent space is not random but based on rules of symmetry, which our WyCryst framework is designed to learn.

| Formula | Space Group | Original Configuration | Equivalent Configuration |
|---|---|---|---|
| $CaTiO_3$ | 62 | {'Ca': ['4c'], 'Ti': ['4b'], 'O': ['4c', '8d']} | {'Ca': ['4c'], 'Ti': ['4a'], 'O': ['4c', '8d']} |
| $CaTiO_3$ | 221 | {'Ca': ['1b'], 'Ti': ['1a'], 'O': ['3d']} | {'Ca': ['1a'], 'Ti': ['1b'], 'O': ['3c']} |
| $SrTiO_3$ | 221 | {'Sr': ['1b'], 'Ti': ['1a'], 'O': ['3d']} | {'Sr': ['1a'], 'Ti': ['1b'], 'O': ['3c']} |
| $Zn_2GeSe_4$ | 122 | {'Zn': ['4d'], 'Ge': ['2b'], 'Se': ['8i']} | {'Zn': ['4d'], 'Ge': ['2a'], 'Se': ['8i']} |

Table 3: Original and equivalent Wyckoff configurations of Wyckoff Genes.

**WyCryst Generation of New Materials**

Using the WyCryst Framework, supported by previous validation results, we now proceed to generate Wykoff Genes for ternary compositions with various total DoFs, which are all exclusive from the training materials database (MP). These were based on a custom element list, which are relatively easy to procure and experimentally accessible to solid-state synthesis labs across the world. The perturbation in the latent space followed a similar principle to validation tasks above, with the initial sampling points were with all existing ternary combinations of these elements in MP. All hyperparameters are tuned based on the leave-one-out validations process to obtain optimal results. To compute the final set of Wyckoff Genes, we applied a series of filters (reliability and feasibility filters) to not only check for the crystal structure fidelity but also exclude materials with low thermodynamic stability (high formation energy) or low synthesizability score. These filters prevent the resources wasted on unrealistic or unstable materials in the subsequent DFT workflow.

Within these constraints, we produced a total of 639 Wyckoff Genes in the generation step, out of which 102 were selected that satisfy the following two criteria: (1) number of atoms per unit cell <= 50; (2) and total DoFs <= 5. These were fed into the automated DFT workflow. Only one relaxed structure, out of total DoF + 1 structures, with the lowest total energy for each Wyckoff Genes was then picked for the final phonon stability check. All the formation energies of DFT-relaxed structures for the selected Wyckoff Genes are negative, consistent with the $E_f$ predicted by our PVAE model. All 102 Wyckoff Genes fed into DFT calculations and the Distribution of crystal systems of 639 Wyckoff Genes are provided in the Supplementary material section 9.

**Structure Analysis of Generated Crystals**

We conducted further detailed analysis of the dynamic stability with respect to phonons of the final Wyckoff Gene, giving rise to 8 new, dynamically stable materials: $Cu_3SnSe_4$, $RbBiTe_2$, $BiGeTe_2$, $Bi_4ZrTe_6$, $Bi_3Ge_4Te_4$, $AgSbTe_4$, $Bi_2TiTe_4$, and $Bi_2VTe_4$. Their structures and phonon dispersions are shown in Figure 4. Among them, only $RbBiTe_2$ (2) was claimed to be successfully synthesized in 1978 (but not recorded in the MP database); albeit, rather unfortunately, without structural analysis[51]. In addition, the remaining 7 dynamically stable materials are neither in the MP database nor found in literature. As for the compound $Cu_3SnSe_4$ (1), it is a stannite mineral and is structurally close to an existing stannite $Cu_3SbSe_4$ for thermoelectric applications[52]. They have the same space group symmetry but are represented in two sets of equivalent Wyckoff positions, respectively: (i) Cu (2a, 4d), Sn (2b), Se (8i) for $Cu_3SnSe_4$, (ii) Cu (2b, 4d), Sb (2a), Se (8i) for $Cu_3SbSe_4$. Notably, the phonon dispersion of $Cu_3SnSe_4$, as shown in Figure 4(1), highly resembles that of $Cu_3SbSe_4$ (Figure S6), implying its high probability of being stable experimentally. These results strongly suggest that it's worth to validate the synthesizability of $Cu_3SnSe_4$ in further research efforts.

The literature that reports $Cu_5Sn_2Se_7$[53] also attached a CIF file with a composition of $Cu_{2.8}Sn_{1.2}Se_4$, which has the same space group symmetry (#121, $I\bar{4}2m$) as the predicted $Cu_3SnSe_4$, but shows the cationic partial occupations (compositional/structural disorder). However, due to the absence of any technical details and discussion on $Cu_{2.8}Sn_{1.2}Se_4$ in the study that focuses on the ordered $Cu_5Sn_2Se_7$, the likely existence of cation disordering which would make ordered $Cu_3SnSe_4$ be $Cu_{2.8}Sn_{1.2}Se_4$ should be taken with a grain of salt. From the perspective of inorganic chemistry, though, the ionic disordering is commonly observed for a large variety of materials, such that it significantly affects the expected properties of predicted crystals.[54,55] To date, nevertheless, there are only a few reports on the CSP for disordered materials, whether it is coupled with ML or not. Conventional ways to investigate the disordered crystals are either to chemically enumerate independent structures for ions with partial occupancy, starting from a prototype of crystals (e.g., spinel[56], perovskite[57]) by the supercell method[58] and special quasi-random structure (SQS)[59], which require an enormous amount of computational time, or to adopt the virtual crystal approximation (VCA)[60] in the DFT calculations. Further, definition of phonon modes with compositional disorder, especially in the context of stability, requires further work.[54,61] It means that much effort is required, beyond the scope of our current study, for the adapting a generative model to the CSP task for disordered materials. This might involve the development of a revised ML model that is trained on a sufficiently large, high-fidelity dataset of disordered crystals through an appropriate crystal representation, and a set of reliable metrics of stability evaluation based on either DFT calculations or ML itself. Including such efforts within our representation is conceivable and our framework is robust to tackle the ground-state and polymorph CSP tasks as well as the properties-directed material generation task.

The $BiGeTe_2$ (3), structurally similar to the O3-type oxide[62], yields a combination of oxidation states of +2 for both Bi and Ge, and −2 for Te, given that Bi can have unusual oxidation states of −3 ($Mg_3Bi_2$), −2 (oxybismuthide), +1 (BiCl), +2 (bismuth carboxylate), and +5 ($BiF_5$) even though it is dominated by +3 state[63]. The Bi(II) also exist in the predicted compound $Bi_4ZrTe_6$ (4). The unusual oxidation state is not exclusive to Bi-based compounds but to a wide range of elements, such as In (I) in the $ABX_2$-type InTe[64,65], Ag(III) in $AgTe_3$[66,67]. And more than that, the oxidation states for some minerals (the shandite $Ni_3Pb_2S_2$[68]) and layered metal-tellurides ($Fe_3GeTe_2$ and $Ni_3GeTe_2$[69] as well as $Fe_3GaTe_2$[70]) can not be trivially explained using the conventional inorganic chemistry. The predicted crystal $Bi_3Ge_4Te_4$ (5) could be represented by such non-trivial case. As for the $AgSbTe_4$ (6), both Ag (III) and Sb (V) adopt possible, albeit the unusual oxidation states. While, as plotted in Figure S8, the calculated $E_{hull}$ of 0.38 eV/atom indicates the relatively weak stability compared with the competing stable binary phases (AgTe, $AgTe_3$ and $SbTe_2$). The WyCryst framework, which adopts SMACT[71] as the test strategy for charge neutrality, allows the unusual oxidation state (or unusual compositions) to be present in the generated crystals, generalizing the capability of crystal structure generation in the inverse design of new materials from the yet-to-explore chemical spaces.

The compositions and the space group symmetry of $Bi_2TiTe_4$ (7) and $Bi_2VTe_4$ (8) might be leant by our PVAE model from the existence of the prototype structure $Bi_2MnTe_4$[72,73] in the training data, where the Bi is in the regular +3 oxidation state and Ti, V, and Mn are in +2 state (all of them are accessible according to the Shannon's ionic radii table[74]). Moreover, they share the same set of Wyckoff positions: Ti/V/Mn at 3a, Bi and Te at 4c sites. One can find in Figure 4 that the Bi-Bi interatomic distances are relatively short (3.0-3.5 Å) and are close to two times the covalent atomic radius of Bi (1.6 Å[75]). This is not so common to observed in the Bi-based inorganic crystals even though, theoretically, the short metal-metal bonds are possible, for example, in the metal monochalcogenide GaS (with Ga-Ga bond length of 2.4 Å)[76]. Such predictions could be improved by introducing the feature of coordination environment into the ML model in the future improvements.

Our results strongly demonstrate the capability of the WyCryst framework to generate the high-validity, symmetry-compliant new crystal structures, and even to produce dynamic (phonon) stable novel materials. Nevertheless, with careful scrutiny one can find that there are two atomic layers of closely-positioned Bi atoms in $Bi_4ZrTe_6$, $Bi_2TiTe_4$ and $Bi_2VTe_4$, possibly implying that they are not likely to be experimentally stable at finite temperature because of their unusual bonding and coordination as expected in typical inorganic crystals. The unusual bonding would be excluded if the model was trained with the information of coordination environments of elements, but this is beyond the scope of our current work. This also implies that the phonon dispersions obtained from the harmonic approximation at 0 K is not good enough to predict the stability of the given crystal structures and is certainly an area of improvement for future work.

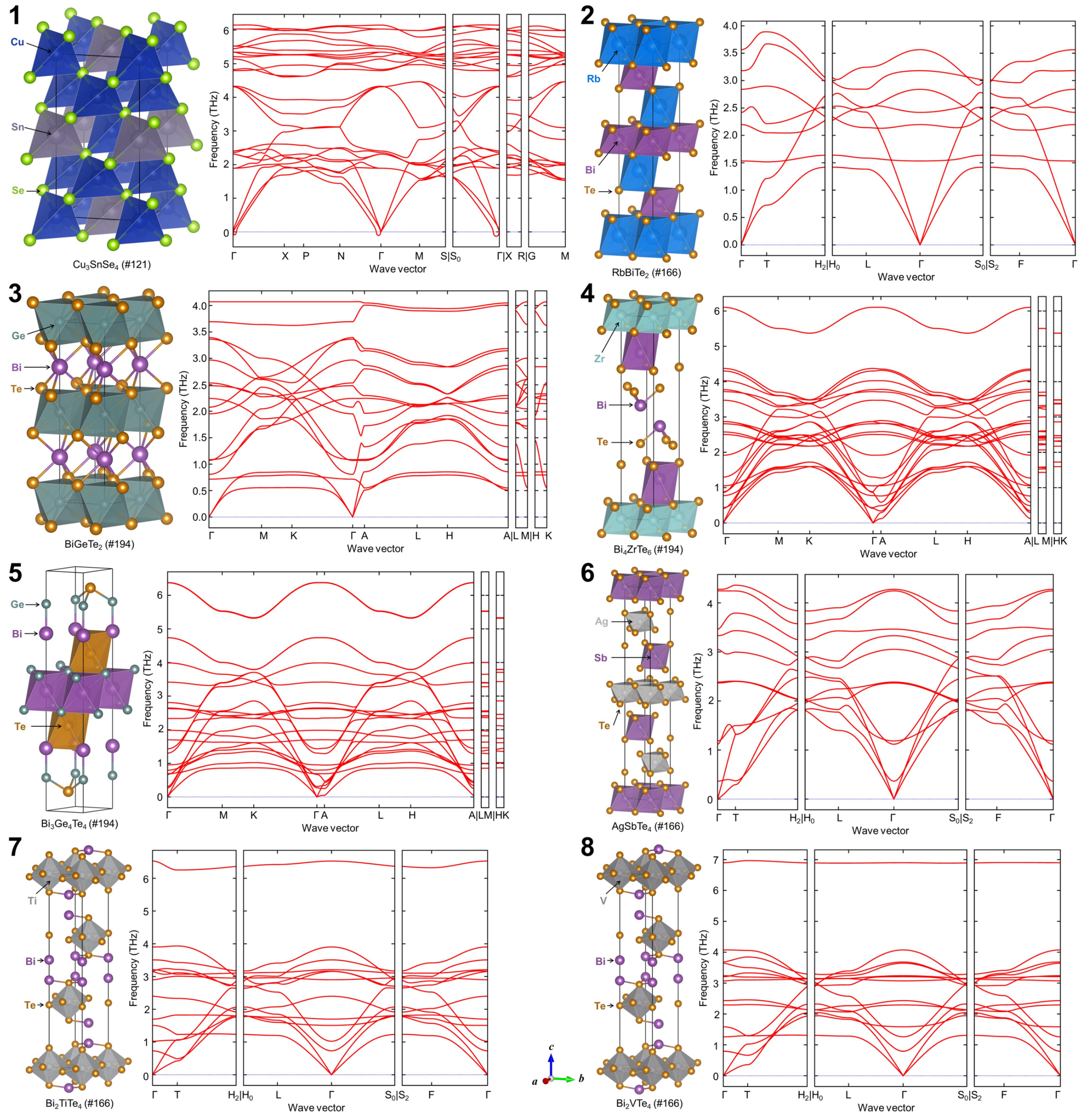

Figure 4: Structures and phonon dispersions of 8 phonon-stable new crystals (not found in the MP database)

# Conclusions

In conclusion, addressing the grand challenge of CSP in inorganic crystalline materials necessitates a more efficient and physics-aware approach that incorporates both the strengths of DFT and also leverages the capabilities of data-driven machine learning. Symmetry considerations, being fundamental to crystal structures, must be seamlessly integrated into any generative design and feature representation. Our WyCryst framework builds on Wyckoff position-based crystal representation that explicitly accounts for such considerations and handles computational tractable DFT energy relaxation and structure refinement. WyCryst not only predicts the ground state crystal structure, but also other structures with the same composition, known as polymorphs. Our Wyckoff-based forward model has demonstrated performance surpassing current state-of-the-art methodologies. Through rigorous validation, our approach addresses both ground state and polymorph tasks, leading to the generation of 8 novel ternary compounds, further validated by checks on phonon stability.

## DATA AVAILABILITY

The entire dataset for training, testing and validation are queried from the Materials Project database (v2023.7.4). Source code for the PVAE model, Wyckoff Gene post-processing, and crystal structure generation is available at https://github.com/RaymondZhurm/WyCryst.

## ACKNOWLEDGEMENTS

K.H. acknowledges funding from the Accelerated Materials Development for Manufacturing Pro- gram at A*STAR via the AME Programmatic Fund by the Agency for Science, Technology and Research under Grant No. A1898b0043. K.H. also acknowledges funding from the NRF Fellowship NRF-NRFF13-2021-0011. We acknowledge Siyu Isaac Parker Tian for his intellectual contributions toward WyCryst model optimization.

## AUTHOR CONTRIBUTIONS

K.H. conceived the research. R.Z. worked closely with W.N. and S.Y. and built and refined the Wyckoff representation and PVAE model. W.N. and R.Z. developed the automatic DFT workflow for crystal structure refinement. K.H., R.Z., W.N. wrote the manuscript, with input from all co-authors.

## COMPETING INTERESTS

K.H. owns equity in a startup focused on using machine learning for materials discovery.

Supplementary material for

**WyCryst: Wyckoff Inorganic Crystal Generator Framework**

**Ruiming Zhu[1,2], Wei Nong[1], Shuya Yamazaki[1,2], Kedar Hippalgaonkar[1,2*]**

**[1] School of Materials Science and Engineering, Nanyang Technological University, Singapore 639798, Singapore**

**[2] Institute of Materials Research and Engineering, Agency for Science, Technology and Research (A*STAR), Singapore 138634, Singapore**

**Correspondence to: kedar@ntu.edu.sg**

## 1. Density functional theory calculations

For the automated density functional theory (DFT) workflow, structure refinement and total energy calculations, density functional perturbation theory (DFPT) were performed using the Vienna ab-initio simulation package (VASP) with the plane-wave basis set[1]. The electron-ion interaction is described by the projector augmented wave (PAW) pseudo-potentials[2]. The exchange-correlation of valence electrons is described using the Perdew-Burke-Ernzerhof (PBE) functional within the generalized gradient approximation (GGA) [3]. The calculations were done using two settings (normal and precise settings) for the energy cutoffs of the plane wave basis, *k*-point meshes, convergence criteria of total energies and forces. ***Normal setting***: a default of energy cutoff (the largest ENMAX parameter on the pseudo-potentials files), a Γ-centered k-mesh with *k*-spacing of 0.2 $Å^{-1}$, $10^{-6}$ eV for total energy, and $10^{-3}$ eV $Å^{-1}$ per atom for force were used. ***Precise setting***: an energy cutoff 520 eV, a Γ-centered k-mesh with *k*-spacing of 0.15 $Å^{-1}$, $10^{-8}$ eV for total energy, and $10^{-4}$ eV $Å^{-1}$ per atom for force were used. For leave-one-out validation, the normal setting was used for structure refinements, while for the new generated crystals, the normal setting was picked first then followed by a precise setting for structure refinements. The DFPT calculations always adopted the precise setting. The tetrahedron method with Blöchl corrections is employed for orbital occupancy[4]. The Monkhorst-Pack scheme is used to sample *k*-points in the Brillouin-zone[5]. The atomic positions are relaxed by the conjugate gradient algorithm. The space group symmetry of the input structure is used, as it is by default, for the symmetrization of charge density with the default threshold for symmetry detection (ISYM = 2 and SYMPREC = $10^{-5}$).

## 2. Hyperparameter tuning in PCA

The hyperparameters for forward models, PVAE models, leave-one-out validation models, and material generation models are tuned using Grid Search Cross-Validation method. For each model, performance of all combinations of tunable hyperparameters are evaluated to determine the optimal set. Tunable hyperparameters of forward models and PVAE models include learning_rate in [4e-4, 2e-4,1e-4, 4e-5, 2e-5, 1e-5], batch_size in [64, 128, 256, 512], and epochs in [64, 72, 128, 256]. One the other hand, since the leave-one-out models and generation models employ the same model structure and same dataset (before the leave-one-out operation), one set of hyperparameters are tuned for all these models based on the entire dataset. These tunable hyperparameters include latent_space_dimension ranging in [64, 128, 256, 512], learning_rate in [4e-4, 2e-4,1e-4, 4e-5, 2e-5, 1e-5], batch_size in [64, 128, 256, 512], and epochs in [64, 72, 128, 256]. In the generation model, the coefficient of local perturbation is tuned independently, ranging in [0.2, 0.4, 0.6 0.8, 1.0, 1.2, 1.4, 1.6, 2.0, 2.5, 3.0]. Final sets of parameters used in these models are provided in the source code.

**3. Machine-learned interatomic potential**

Table S1. The lattice parameters and atomic positions of $CaTiO_3$ generated by PyXtal as the initial unput structure and the subsequent relaxed values after relaxation

| | |
|---|---|
| Initial input structure | $CaTiO_3$, *Pnma* (62)<br>*a* = 6.851405 Å<br>*b* = 10.796344 Å<br>*c* = 4.995430 Å<br>*α* = *β* = *γ* = 90°<br><br>Ca: 4c (0.231906, 0.25, 0.242843)<br>Ti: 4b (0, 0, 0.5)<br>O1: 4c, (0.530608, 0.25, 0.830432)<br>O2: 8d (0.504495, 0.343034, 0.463779 |
| M3GNET relaxed (without symmetry constraint for structure relaxation) | $CaTiO_3$, $P2_1/c$ (14)<br>*a* = 6.090907<br>*b* = 17.136172<br>*c* = 3.120845<br>*α* = *β* = 90°, *γ* = 53.128434<br><br>Ca: 4e (0.260953, 0.249228, 0.249636)<br>Ti: 2b (0, 0, 0.5)<br>Ti: 2a (0, 0.5, 0.5)<br>O1: 4e (0.508942, 0.250079, 0.744699)<br>O2: 4e (0.469723, 0.431506, 0.400853)<br>O3: 4e (0.467564, 0.931813, 0.598798) |
| M3GNET force-diverged* (impose symmetry constraint for structure relaxation)<br><br><br><br><br>*The last trajectory with max force of 0.508 eV $Å^{-1}$ per atom after 800 ionic steps | $CaTiO_3$, *Pnma* (62)<br>*a* = 6.20816<br>*b* = 14.50380<br>*c* = 3.08004<br>*α* = *β* =*γ* = 90°<br><br>Ca: 4c (0.32204, 0.25, 0.25306)<br>Ti: 4b (0, 0, 0.5)<br>O1: 4c (0.57299, 0.25, 0.75461)<br>O2: 8d (0.35992, 0.43996, 0.44004) |
| DFT(VASP) relaxed | $CaTiO_3$, *Pnma* (62)<br>*a* = 5.506974 Å<br>*b* = 7.688278 Å<br>*c* = 5.395221 Å<br>*α* = *β* = *γ* = 90°<br><br>Ca (4c): (0.957402, 0.25, −0.009446)<br>Ti (4b): (0.5, 0.5, 0)<br>O1 (4c): (−0.020181, 0.75, 0.578285)<br>O2 (8d): (0.210449, 0.458873, −0.210068) |

**4. Computational time**

Table S2. Summary of the average computation times of the WyCryst framework

| Step | Computational time |
|---|---|
| Model training | 2 GPU seconds/iteration |
| Latent space sampling | 1 CPU second/2000 structures |
| PyXtal generation | 2 CPU core seconds/structure |
| DFT relaxation | 110 CPU core hours/structure |
| Phonon calculations | 147 CPU core hours/structure |

## 5. Reconstructed Wyckoff Genes

Table S3. The Wyckoff Genes for ground-state $CaTiO_3$ CSP sorted by DoF

| # | Reconstructed formula | Wyckoff site occupancy | SG No. | Predicted $E_f$ (eV/atom)[a] | DoF[b] |
|---|---|---|---|---|---|
| 0 | $Ca_4Ti_4O_{12}$ | {'Ca': ['4b'], 'Ti': ['4c'], 'O': ['4a', '8e']} | 140 | −3.528 | 0 |
| 1 | $Ca_4Ti_4O_{12}$ | {'Ca': ['4b'], 'Ti': ['4c'], 'O': ['4a', '8g']} | 140 | −3.566 | 1 |
| 2 | $Ca_4Ti_4O_{12}$ | {'Ca': ['4b'], 'Ti': ['4d'], 'O': ['4a', '8h']} | 140 | −3.564 | 1 |
| 3 | $Ca_4Ti_4O_{12}$ | {'Ca': ['4b'], 'Ti': ['4d'], 'O': ['4a', '8f']} | 140 | −3.563 | 1 |
| 4 | $Ca_4Ti_4O_{12}$ | {'Ca': ['4b'], 'Ti': ['4c'], 'O': ['4d', '8h']} | 140 | −3.563 | 1 |
| 5 | $Ca_4Ti_4O_{12}$ | {'Ca': ['4b'], 'Ti': ['4c'], 'O': ['4a', '8f']} | 140 | −3.562 | 1 |
| 6 | $Ca_4Ti_4O_{12}$ | {'Ca': ['4b'], 'Ti': ['4c'], 'O': ['4a', '8h']} | 140 | −3.549 | 1 |
| 7 | $Ca_4Ti_4O_{12}$ | {'Ca': ['4b'], 'Ti': ['4a'], 'O': ['4c', '8h']} | 140 | −3.487 | 1 |
| 8 | $Ca_4Ti_4O_{12}$ | {'Ca': ['4e'], 'Ti': ['4b'], 'O': ['4a', '4d', '4e']} | 74 | −3.631 | 2 |
| 9 | $Ca_4Ti_4O_{12}$ | {'Ca': ['4e'], 'Ti': ['4b'], 'O': ['4a', '8f']} | 74 | −3.592 | 2 |
| 1 | $Ca_4Ti_4O_{12}$ | {'Ca': ['4e'], 'Ti': ['4c'], 'O': ['4d', '8g']} | 74 | −3.578 | 2 |
| 1 | $Ca_4Ti_4O_{12}$ | {'Ca': ['4e'], 'Ti': ['4a'], 'O': ['4d', '8g']} | 74 | −3.569 | 2 |
| 1 | $Ca_4Ti_4O_{12}$ | {'Ca': ['4e'], 'Ti': ['4b'], 'O': ['4a', '8g']} | 74 | −3.565 | 2 |
| 1 | $Ca_4Ti_4O_{12}$ | {'Ca': ['4e'], 'Ti': ['4b'], 'O': ['4d', '8g']} | 74 | −3.555 | 2 |
| 1 | $Ca_4Ti_4O_{12}$ | {'Ca': ['4e'], 'Ti': ['4b'], 'O': ['4c', '4d', '4e']} | 74 | −3.554 | 2 |
| 1 | $Ca_4Ti_4O_{12}$ | {'Ca': ['4e'], 'Ti': ['4b'], 'O': ['4c', '8f']} | 74 | −3.549 | 2 |
| 1 | $Ca_4Ti_4O_{12}$ | {'Ca': ['4a'], 'Ti': ['4b'], 'O': ['4e', '8g']} | 74 | −3.522 | 2 |
| 1 | $Ca_4Ti_4O_{12}$ | {'Ca': ['4e'], 'Ti': ['4b'], 'O': ['4c', '8g']} | 74 | −3.517 | 2 |
| 1 | $Ca_4Ti_4O_{12}$ | {'Ca': ['4c'], 'Ti': ['4b'], 'O': ['4e', '8g']} | 74 | −3.494 | 2 |
| 1 | $Ca_4Ti_4O_{12}$ | {'Ca': ['4e'], 'Ti': ['4c'], 'O': ['4b', '8g']} | 74 | −3.491 | 2 |
| 2 | $Ca_4Ti_4O_{12}$ | {'Ca': ['4d'], 'Ti': ['4b'], 'O': ['4e', '8g']} | 74 | −3.480 | 2 |
| 2 | $Ca_4Ti_4O_{12}$ | {'Ca': ['4e'], 'Ti': ['4a'], 'O': ['4e', '8g']} | 74 | −3.600 | 3 |
| 2 | $Ca_4Ti_4O_{12}$ | {'Ca': ['4e'], 'Ti': ['4b'], 'O': ['4a', '8h']} | 74 | −3.570 | 3 |
| 2 | $Ca_4Ti_4O_{12}$ | {'Ca': ['4e'], 'Ti': ['4b'], 'O': ['4e', '8g']} | 74 | −3.562 | 3 |
| 2 | $Ca_4Ti_4O_{12}$ | {'Ca': ['4e'], 'Ti': ['4b'], 'O': ['4e', '8f']} | 74 | −3.562 | 3 |
| 2 | $Ca_4Ti_4O_{12}$ | {'Ca': ['4e'], 'Ti': ['4c'], 'O': ['4e', '8g']} | 74 | −3.548 | 3 |
| 2 | $Ca_4Ti_4O_{12}$ | {'Ca': ['4e'], 'Ti': ['4e'], 'O': ['4c', '8g']} | 74 | −3.506 | 3 |
| 2 | $Ca_4Ti_4O_{12}$ | {'Ca': ['4e'], 'Ti': ['4d'], 'O': ['4e', '8g']} | 74 | −3.475 | 3 |
| 2 | $Ca_4Ti_4O_{12}$ | {'Ca': ['4e'], 'Ti': ['4e'], 'O': ['4e', '8g']} | 74 | −3.576 | 4 |
| 2 | $Ca_4Ti_4O_{12}$ | {'Ca': ['4c'], 'Ti': ['4c'], 'O': ['4a', '8g']} | 63 | −3.536 | 4 |
| 3 | $Ca_4Ti_4O_{12}$ | {'Ca': ['4e'], 'Ti': ['4b'], 'O': ['4e', '8h']} | 74 | −3.532 | 4 |
| 3 | $Ca_4Ti_4O_{12}$ | {'Ca': ['4c'], 'Ti': ['4a'], 'O': ['4c', '8f']} | 63 | −3.414 | 4 |
| 3 | $Ca_4Ti_4O_{12}$ | {'Ca': ['4c'], 'Ti': ['4b'], 'O': ['4c', '8f']} | 63 | −3.410 | 4 |
| 3 | $Ca_4Ti_4O_{12}$ | {'Ca': ['4c'], 'Ti': ['4b'], 'O': ['4c', '8g']} | 63 | −3.372 | 4 |
| 3 | $Ca_4Ti_4O_{12}$ | {'Ca': ['4b'], 'Ti': ['4a'], 'O': ['4c', '8d']} | 62 | −3.533 | 5 |
| 3 | $Ca_4Ti_4O_{12}$ | {'Ca': ['4e'], 'Ti': ['4a'], 'O': ['4e', '8f']} | 15 | −3.492 | 5 |
| 3 | $Ca_4Ti_4O_{12}$ | {'Ca': ['4c'], 'Ti': ['4b'], 'O': ['4c', '8d']} | 62 | −3.546 | 7 |

[a] Values that were sampling from our PVAE model.

[b] Counting the total DoF of a Wyckoff Gene from that of Wyckoff positions, taking the # 36 as an example: the Ca in 4c (*x*, 1/4, *z*) contributes two DoF, the Ti in 4b (0, 0, 1/2) contribute zero DoF, and O in 4c (*x*, 1/4, *z*) and 8d (*x*, *y*, *z*) gives rise to two and three DoF, respectively. So, we have the DoF of this Wyckoff Gene is 2 + 0 + 2 + 3 = 7. The symmetry-dependent Wyckoff positions can refer to the International Tables for Crystallography.

Table S4. The Wyckoff Genes for ground-state $BaTiO_3$ CSP sorted by DoF

| # | Reconstructed formula | Wyckoff site occupancy | SG No. | Predicted $E_f$ (eV/atom) | DoF |
|---|---|---|---|---|---|
| 0 | $BaTiO_3$ | {'Ba': ['1b'], 'Ti': ['1a'], 'O': ['3c']} | 221 | −3.430 | 0 |
| 1 | $Ba_4Ti_4O_{12}$ | {'Ba': ['4a'], 'Ti': ['4b'], 'O': ['4c', '8e']} | 140 | −3.428 | 0 |
| 2 | $BaTiO_3$ | {'Ba': ['1b'], 'Ti': ['1a'], 'O': ['3d']} | 221 | −3.424 | 0 |
| 3 | $Ba_2Ti_2O_6$ | {'Ba': ['2b'], 'Ti': ['2c'], 'O': ['6h']} | 194 | −3.434 | 1 |
| 4 | $Ba_4Ti_4O_{12}$ | {'Ba': ['4b'], 'Ti': ['4a'], 'O': ['4c', '8g']} | 74 | −3.408 | 1 |
| 5 | $Ba_2Ti_2O_6$ | {'Ba': ['2b'], 'Ti': ['2a'], 'O': ['2c', '4g']} | 65 | −3.330 | 1 |
| 6 | $Ba_4Ti_4O_{12}$ | {'Ba': ['4c'], 'Ti': ['4a'], 'O': ['4b', '8h']} | 74 | −3.360 | 2 |
| 7 | $Ba_4Ti_4O_{12}$ | {'Ba': ['4e'], 'Ti': ['2a', '2d'], 'O': ['4f', '8j']} | 12 | −3.505 | 3 |
| 8 | $Ba_4Ti_4O_{12}$ | {'Ba': ['4g'], 'Ti': ['2a', '2d'], 'O': ['4f', '8j']} | 12 | −3.556 | 4 |
| 9 | $Ba_4Ti_4O_{12}$ | {'Ba': ['4a'], 'Ti': ['4a'], 'O': ['4a', '8b']} | 42 | −3.451 | 4 |
| 10 | $BaTiO_3$ | {'Ba': ['1b'], 'Ti': ['1a'], 'O': ['1a', '2c']} | 99 | −3.414 | 4 |
| 11 | $BaTiO_3$ | {'Ba': ['1a'], 'Ti': ['1a'], 'O': ['1a', '2c']} | 99 | −3.385 | 4 |
| 12 | $Ba_3Ti_3O_9$ | {'Ba': ['3a'], 'Ti': ['3a'], 'O': ['9b']} | 160 | −3.242 | 4 |
| 13 | $Ba_4Ti_4O_{12}$ | {'Ba': ['4e'], 'Ti': ['4i'], 'O': ['4f', '8j']} | 12 | −3.550 | 5 |
| 14 | $Ba_2Ti_2O_6$ | {'Ba': ['2a'], 'Ti': ['2a'], 'O': ['2a', '4d']} | 38 | −3.517 | 5 |
| 15 | $Ba_2Ti_2O_6$ | {'Ba': ['2a'], 'Ti': ['2a'], 'O': ['2a', '4e']} | 38 | −3.502 | 5 |
| 16 | $Ba_2Ti_2O_6$ | {'Ba': ['2a'], 'Ti': ['2b'], 'O': ['2a', '4c']} | 38 | −3.500 | 5 |
| 17 | $Ba_2Ti_2O_6$ | {'Ba': ['2a'], 'Ti': ['2b'], 'O': ['2a', '4e']} | 38 | −3.486 | 5 |
| 18 | $Ba_2Ti_2O_6$ | {'Ba': ['2b'], 'Ti': ['2b'], 'O': ['2a', '4e']} | 38 | −3.479 | 5 |
| 19 | $Ba_2Ti_2O_6$ | {'Ba': ['2a'], 'Ti': ['2b'], 'O': ['2a', '4d']} | 38 | −3.473 | 5 |
| 20 | $Ba_2Ti_2O_6$ | {'Ba': ['2a'], 'Ti': ['2b'], 'O': ['2a', '2a', '2b']} | 38 | −3.440 | 5 |
| 21 | $Ba_4Ti_4O_{12}$ | {'Ba': ['4b'], 'Ti': ['4a'], 'O': ['4c', '8d']} | 62 | −3.438 | 5 |
| 22 | $Ba_2Ti_2O_6$ | {'Ba': ['2a'], 'Ti': ['2b'], 'O': ['2b', '4e']} | 38 | −3.402 | 5 |
| 23 | $Ba_2Ti_2O_6$ | {'Ba': ['2a'], 'Ti': ['2a'], 'O': ['2b', '4e']} | 38 | −3.370 | 5 |
| 24 | $Ba_4Ti_4O_{12}$ | {'Ba': ['4b'], 'Ti': ['4c'], 'O': ['4c', '8d']} | 62 | −3.437 | 7 |
| 25 | $Ba_4Ti_4O_{12}$ | {'Ba': ['4c'], 'Ti': ['4e'], 'O': ['4c', '4d', '4e']} | 38 | −3.460 | 10 |

Table S5. Crystallographic data for ground-state $CaTiO_3$ and $BaTiO_3$

| | $CaTiO_3$ | | $BaTiO_3$ | |
|---|---|---|---|---|
| | experimental | DFT-refined | experimental | DFT-refined |
| System | orthorhombic | orthorhombic | trigonal* | trigonal* |
| SG No. | 62 | 62 | 160 | 160 |
| Lattice parameters | $a$ = 5.44 Å<br>$b$ = 7.64 Å<br>$c$ = 5.37 Å<br>$\alpha = \beta = \gamma$ = 90° | $a$ = 5.51 Å<br>$b$ = 7.69 Å<br>$c$ = 5.37 Å<br>$\alpha = \beta = \gamma$ = 90° | $a = b = c$ = 4.00 Å<br>$\alpha = \beta = \gamma$ = 89.8° | $a = b = c$ = 4.07 Å<br>$\alpha = \beta = \gamma$ = 89.7° |
| Wyckoff positions | Ca (4c): (0.04, 3/4, 0.99)<br>Ti: (4b): (0, 0, 1/2)<br>O (4c): (0.02, 1/4, 0.58)<br>O (8d): (0.21, 0.04, 0.21) | Ca (4c): (0.96, 1/4, 0.01) †<br>Ti: (4b): (0, 0, 1/2)<br>O (4c): (0.02, 1/4, 0.42)<br>O (8d): (0.21, 0.46, −0.21) | Ba (3a): (0, 0, 0.0002)<br>Ti (3a): (0, 0, 0.51)<br>O (9b): (0.17, 0.34, 0.32) | Ba (3a): (0, 0, 0.17) §<br>Ti (3a): (0, 0, 0.68)<br>O (9b): (0.17, 0.34, 0.49) |

* The trigonal crystal system can be represented in both rhombohedral axes (primitive cell) and in hexagonal axes (conventional cell). In our context for Wyckoff Genes and visualization, the hexagonal axes are used, while the rhombohedral axes are adopted when the lattice parameters are stated in the main text, for the sake of differentiation from the hexagonal crystal system. The transformation from rhombohedral axes $(a_R, b_R, c_R)^T$ to hexagonal axes $(a_H, b_H, c_H)^T$ is given by

$$\begin{pmatrix} a_H \\ b_H \\ c_H \end{pmatrix} = \begin{pmatrix} -1 & 1 & 0 \\ 1 & 0 & -1 \\ 1 & 1 & 1 \end{pmatrix} \begin{pmatrix} a_R \\ b_R \\ c_R \end{pmatrix}.$$

† For 4c and 8d coordinates of the DFT-refined structures, they are symmetrically equivalent to the respective values of the experimental one.

§ All coordinates are translated by ~ 0.17 along the $c$ axis.

Table S6. The Wyckoff Genes for polymorphic CSP of $CaTiO_3$, $SrTiO_3$, $CsPbI_3$, and $CuInS_2$

| # | Reconstructed formula | Wyckoff site occupancy | SG No. | Predicted $E_f$ (eV/atom) | DoF |
|---|---|---|---|---|---|
| 0 | $CaTiO_3$ | {'Ca': ['1b'], 'Ti': ['1a'], 'O': ['3d']} | 221 | −3.442 | 0 |
| 1 | $CaTiO_3$ | {'Ca': ['1a'], 'Ti': ['1b'], 'O': ['3c']} | 221 | −3.496 | 0 |
| 2 | $SrTiO_3$ | {'Sr': ['1b'], 'Ti': ['1a'], 'O': ['3d']} | 221 | −3.442 | 0 |
| 3 | $SrTiO_3$ | {'Sr': ['1a'], 'Ti': ['1b'], 'O': ['3c']} | 221 | −3.371 | 0 |
| 4 | $CsPbI_3$ | {'Cs': ['1a'], 'Pb': ['1b'], 'I': ['3c']} | 221 | −1.361 | 0 |
| 5 | $CuInS_2$ | {'Cu': ['4a'], 'In': ['4b'], 'S': ['8d’]} | 122 | −0.684 | 1 |

## 6. Formation energy

The DFT-calculated formation energy, $E_f$, of a refined structure of ternary crystal $A_lB_mC_n$ is calculated according to the following equation,

$$E_f(A_lB_mC_n) = [E_{DFT}(A_lB_mC_n) - (l \times \mu_A + m \times \mu_B + n \times \mu_C)]/(l + m + n) \quad (S1)$$

where $E_{DFT}(A_lB_mC_n)$ is the total energy of the DFT-refined ternary crystal, $\mu_A$, $\mu_B$, and $\mu_C$ are the chemical potentials of element A, B and C, respectively, and $l$, $m$, and $n$ are the composition (content) of the corresponding elements. The chemical potential is adopted as the total energy per atom in the corresponding stable elementary crystalline substance, for example, Ca in face-centered cubic metal (space group No. 225), O in monoclinic molecular crystal of $O_2$ dimmer (space group No. 12).

## 7. Ground-state CSP of rhombohedral $BaTiO_3$

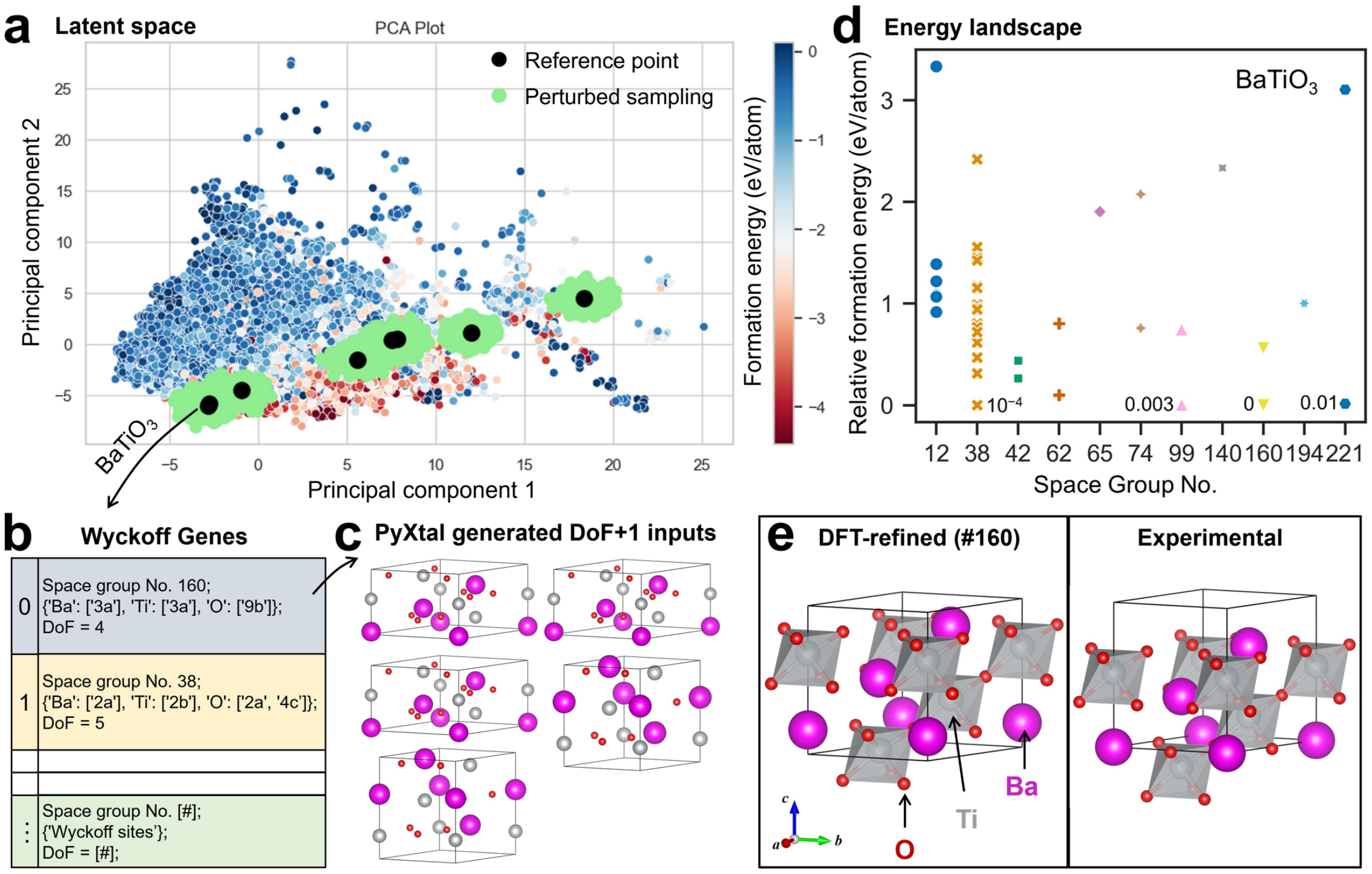

Figure S1. Ground state CSP task including latent space sampling, Wyckoff Genes, and DFT workflow results: (a) PCA visualization of the $BaTiO_3$ (space group No.160) latent space sampling, where black reference points are all materials with the same stoichiometry used as the origin point for sampling and green points are all sampled data points; (b) Wyckoff Genes of sampled materials after post-processing; (c) PyXtal generated crystal structures from the first Wyckoff Gene (space group No. 160); (d) The relative formation energy of all 126 refined $BaTiO_3$ structures; (e) crystal structure comparison between WyCryst generated ground state structure and experimental stable structure reported in literature.

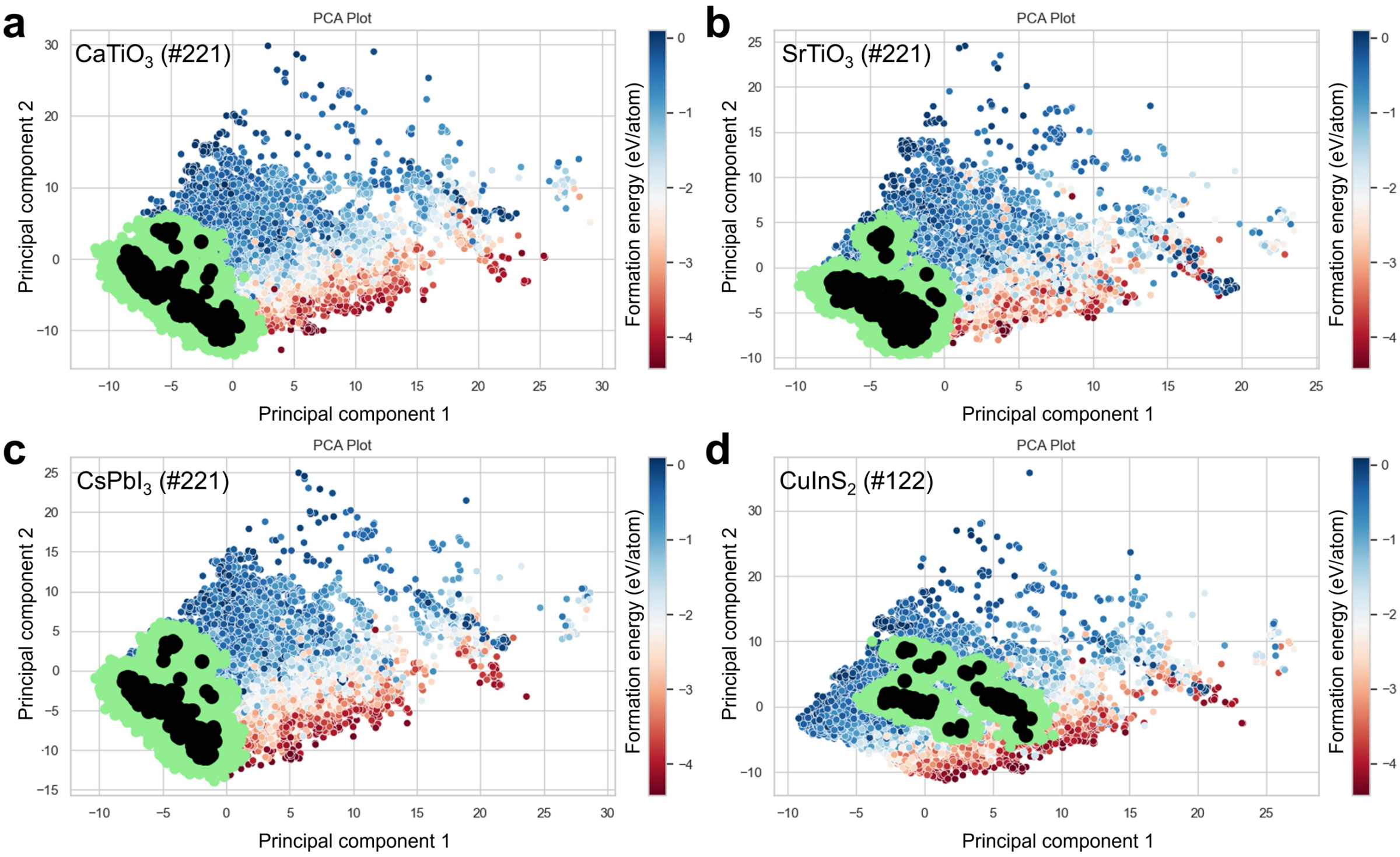

Figure S2. Latent space sampling for polymorph CSP of (a) $CaTiO_3$, (b) SrTiO3, (c) $CsPbI_3$ and (d) $CuInS_2$.

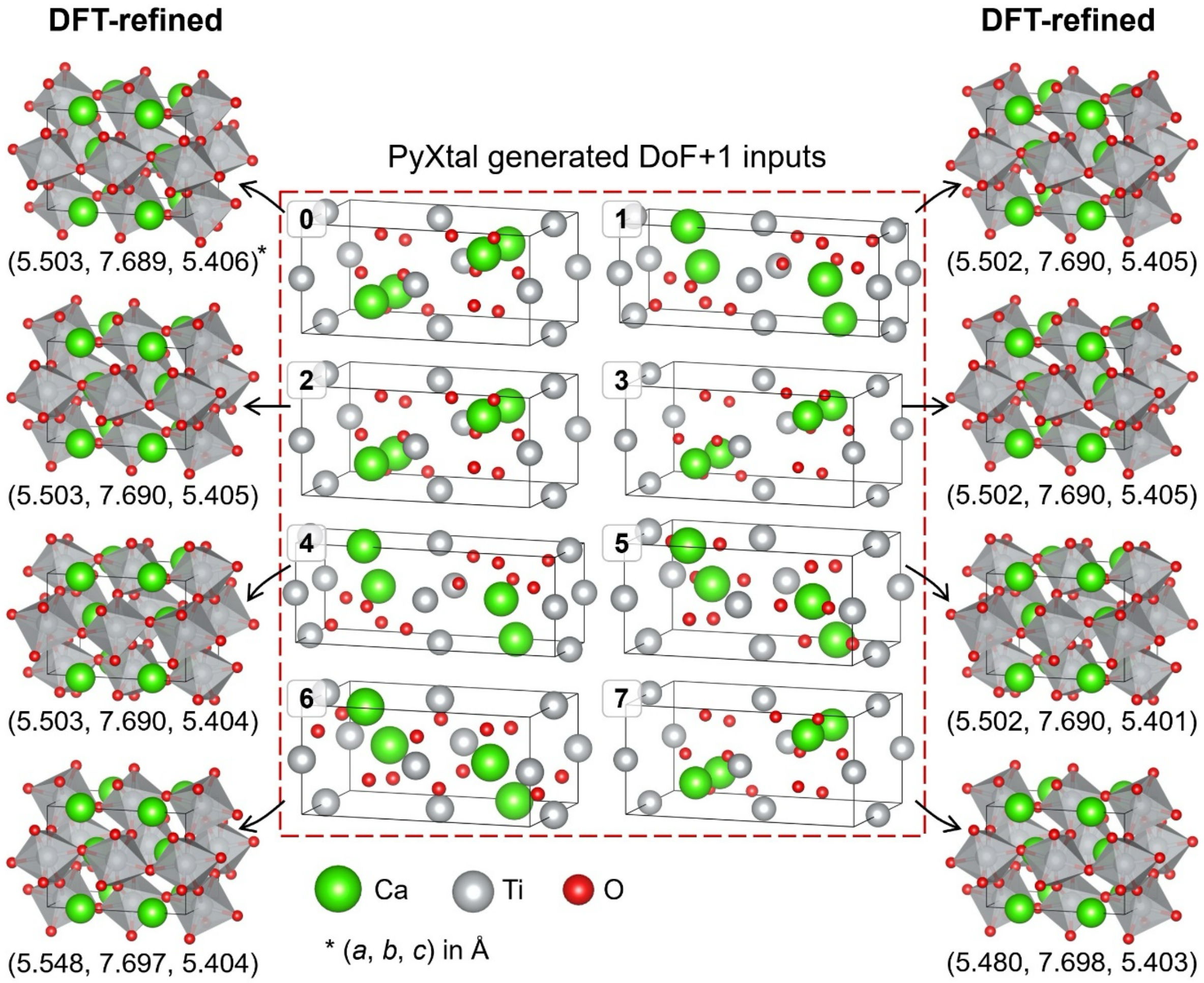

Figure S3. Structural comparison between the initial structures for DFT input and the relaxed structure after DFT relaxation with normal settings for total energy and force. The lowest-energy DFT refined result is given as shown in Figure 2e of the manuscript and Table S5.

## 8. Analysis of WyCryst Training Data from the MP database

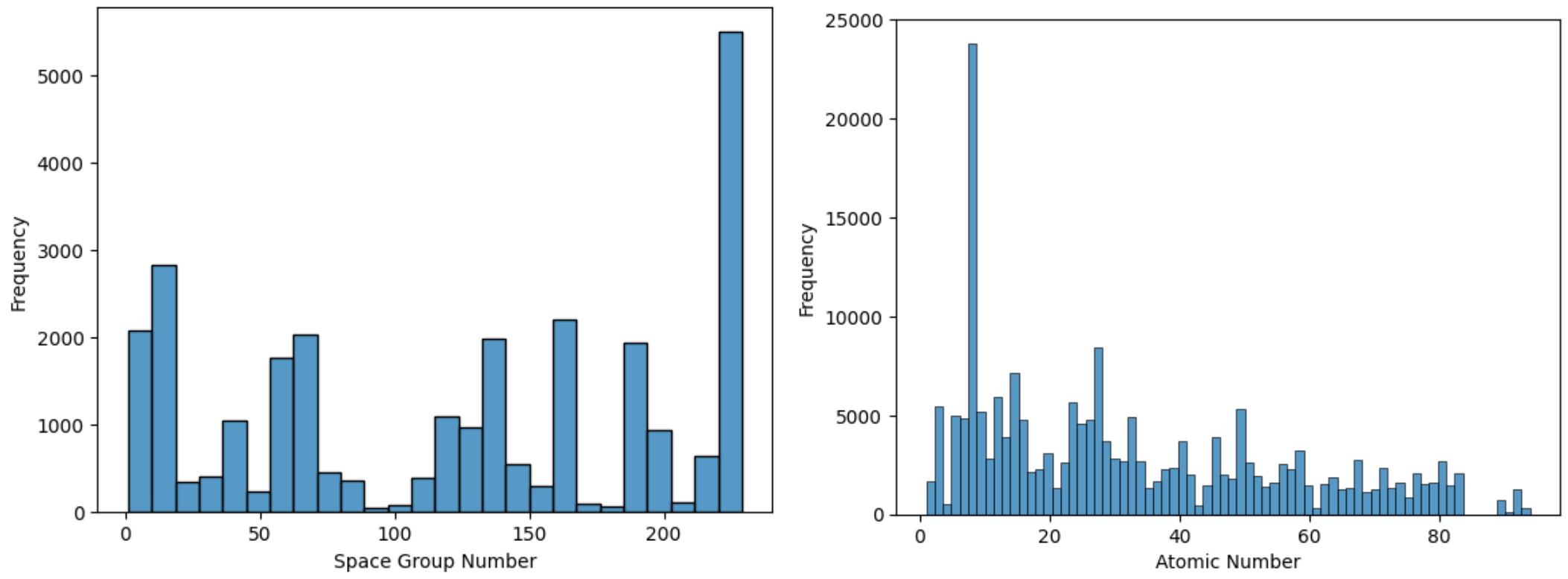

Figure S4. Distribution of space group numbers and atomic numbers of the training set in the MP database.

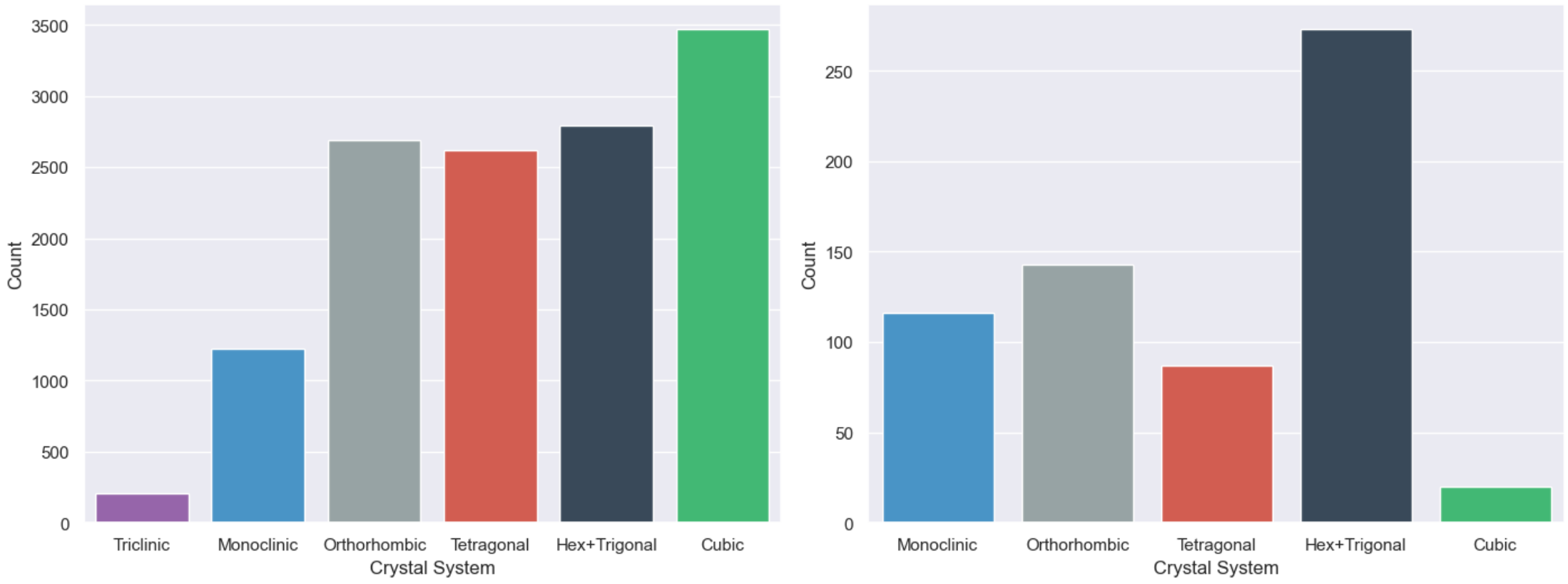

Figure S5. (a) Distribution of crystal systems of materials on the convex hull ($E_{hull}$ = 0 eV/atom) in the MP database, (b) Distribution of crystal systems of 639 Wyckoff Genes (trigonal materials are combined with hexagonal ones because they are presented in hexagonal unit cell for conventional standard setting).

## 9. Wyckoff Genes from a user-defined input

Table S7. The Wyckoff Genes generated the number of atoms per unit cell <= 50 and (2) and total DoFs <= 5.

| # | Reconstructed formula | Wyckoff site occupancy | SG No. | Predicted $E_f$ (eV/atom) | Do | SC score |
|---|---|---|---|---|---|---|
| 0 | Rb3Bi3Te6 | {'Rb': ['3a'], 'Bi': ['3b'], 'Te': ['6c']} | 166 | −0.745 | 1 | 0.50 |
| 1 | Ag3Sb1Te4 | {'Ag': ['3c'], 'Sb': ['1b'], 'Te': ['1a', '3d']} | 221 | −0.254 | 0 | 0.96 |
| 2 | Ag3Sb1Te3 | {'Ag': ['3c'], 'Sb': ['1b'], 'Te': ['3d']} | 221 | −0.248 | 0 | 0.96 |
| 3 | Cu4Ga4Se8 | {'Cu': ['4a'], 'Ga': ['4a'], 'Se': ['4a', '4a']} | 198 | −0.539 | 4 | 0.93 |
| 4 | Cu4Te8Ge4 | {'Cu': ['4a'], 'Te': ['4a', '4a'], 'Ge': ['4a']} | 198 | −0.228 | 4 | 0.54 |
| 5 | Nb2Ge2Se2 | {'Nb': ['2b'], 'Ge': ['2a'], 'Se': ['2a']} | 186 | −0.779 | 3 | 0.79 |
| 6 | Ag3Sn3Te9 | {'Ag': ['3a'], 'Sn': ['3b'], 'Te': ['9e']} | 166 | −0.355 | 0 | 0.65 |
| 7 | Ag3Bi3Te9 | {'Ag': ['3b'], 'Bi': ['3a'], 'Te': ['9e']} | 166 | −0.311 | 0 | 0.66 |
| 8 | Ag3Bi9Te6 | {'Ag': ['3a'], 'Bi': ['9d'], 'Te': ['6c']} | 166 | −0.248 | 1 | 0.90 |
| 9 | Ge3Bi6Te15 | {'Ge': ['3a'], 'Bi': ['6c'], 'Te': ['6c', '9e']} | 166 | −0.488 | 2 | 0.95 |
| 10 | Ag3Bi6Se6 | {'Ag': ['3a'], 'Bi': ['6c'], 'Se': ['6c']} | 166 | −0.470 | 2 | 0.61 |
| 11 | Sn3Sb6Te15 | {'Sn': ['3a'], 'Sb': ['6c'], 'Te': ['6c', '9e']} | 166 | −0.447 | 2 | 0.77 |
| 12 | Si3Bi6Te6 | {'Si': ['3a'], 'Bi': ['6c'], 'Te': ['6c']} | 166 | −0.433 | 2 | 0.98 |
| 13 | Ag6Sn3Se6 | {'Ag': ['6c'], 'Sn': ['3b'], 'Se': ['6c']} | 166 | −0.433 | 2 | 0.74 |
| 14 | Ag18Sn3Se15 | {'Ag': ['18f'], 'Sn': ['3b'], 'Se': ['6c', '9d']} | 166 | −0.285 | 2 | 0.50 |
| 15 | Ge6Bi6Te6 | {'Ge': ['3a', '3b'], 'Bi': ['6c'], 'Te': ['6c']} | 166 | −0.279 | 2 | 0.92 |
| 16 | Ag3Sb3Te12 | {'Ag': ['3b'], 'Sb': ['3a'], 'Te': ['6c', '6c']} | 166 | −0.268 | 2 | 0.51 |
| 17 | Ti3Bi6Te12 | {'Ti': ['3a'], 'Bi': ['6c'], 'Te': ['6c', '6c']} | 166 | −0.802 | 3 | 0.82 |
| 18 | V3Bi6Te12 | {'V': ['3a'], 'Bi': ['6c'], 'Te': ['6c', '6c']} | 166 | −0.643 | 3 | 0.81 |
| 19 | In3Bi6Te21 | {'In': ['3b'], 'Bi': ['6c'], 'Te': ['6c', '6c', '9d']} | 166 | −0.563 | 3 | 0.65 |
| 20 | Sn3Sb6Te12 | {'Sn': ['3b'], 'Sb': ['6c'], 'Te': ['6c', '6c']} | 166 | −0.430 | 3 | 0.63 |
| 21 | Sb9Pb9Te12 | {'Sb': ['9e'], 'Pb': ['3a', '6c'], 'Te': ['6c', '6c']} | 166 | −0.399 | 3 | 0.75 |
| 22 | Ge12Bi9Te12 | {'Ge': ['3a', '9e'], 'Bi': ['3b', '6c'], 'Te': ['6c', '6c']} | 166 | −0.263 | 3 | 0.76 |
| 23 | Ag9Sb6Te15 | {'Ag': ['3b', '6c'], 'Sb': ['6c'], 'Te': ['3a', '6c', '6c']} | 166 | −0.282 | 4 | 0.69 |
| 24 | Ge9Bi9Te12 | {'Ge': ['3a', '6c'], 'Bi': ['3b', '6c'], 'Te': ['6c', '6c']} | 166 | −0.280 | 4 | 0.73 |
| 25 | Ge6Cu12Te15 | {'Ge': ['6c'], 'Cu': ['6c', '6c'], 'Te': ['3b', '6c', '6c']} | 166 | −0.334 | 5 | 0.61 |
| 26 | Na3Ge1Te2 | {'Na': ['1a', '2d'], 'Ge': ['1b'], 'Te': ['2d']} | 164 | −0.611 | 2 | 0.61 |
| 27 | Ag1Sb1I4 | {'Ag': ['1a'], 'Sb': ['1b'], 'I': ['2c', '2d']} | 164 | −0.426 | 2 | 0.75 |
| 28 | Ag1Bi2Te2 | {'Ag': ['1a'], 'Bi': ['2d'], 'Te': ['2d']} | 164 | −0.235 | 2 | 0.95 |
| 29 | Ga3Sn2Te5 | {'Ga': ['3f'], 'Sn': ['2d'], 'Te': ['1a', '2d', '2d']} | 164 | −0.489 | 3 | 0.65 |
| 30 | Ge1Bi2Te4 | {'Ge': ['1a'], 'Bi': ['2d'], 'Te': ['2c', '2d']} | 164 | −0.412 | 3 | 0.95 |
| 31 | Rh4Ge1Te3 | {'Rh': ['2d', '2d'], 'Ge': ['1b'], 'Te': ['1a', '2d']} | 164 | −0.397 | 3 | 0.63 |
| 32 | Ge2Bi2Te4 | {'Ge': ['1a', '1b'], 'Bi': ['2c'], 'Te': ['2d', '2d']} | 164 | −0.387 | 3 | 0.87 |
| 33 | Ag2Bi1Te5 | {'Ag': ['2d'], 'Bi': ['1b'], 'Te': ['1a', '2d', '2d']} | 164 | −0.292 | 3 | 0.58 |
| 34 | Ge3Bi6Te5 | {'Ge': ['3f'], 'Bi': ['1a', '2d', '3e'], 'Te': ['1b', '2c', '2d']} | 164 | −0.290 | 3 | 0.96 |
| 35 | Ag1Bi3Te4 | {'Ag': ['1b'], 'Bi': ['1a', '2d'], 'Te': ['2d', '2d']} | 164 | −0.266 | 3 | 0.88 |
| 36 | Tm2Bi3Te5 | {'Tm': ['2d'], 'Bi': ['1b', '2c'], 'Te': ['1a', '2d', '2d']} | 164 | −1.115 | 4 | 0.62 |
| 37 | Sn1Bi4Te5 | {'Sn': ['1a'], 'Bi': ['2d', '2d'], 'Te': ['1b', '2c', '2d']} | 164 | −0.477 | 4 | 0.51 |
| 38 | In2Bi2Te4 | {'In': ['2d'], 'Bi': ['2c'], 'Te': ['2d', '2d']} | 164 | −0.435 | 4 | 0.58 |
| 39 | Ge2Bi2Te5 | {'Ge': ['2d'], 'Bi': ['2c'], 'Te': ['1a', '2d', '2d']} | 164 | −0.429 | 4 | 0.92 |
| 40 | Ge3Bi2Te5 | {'Ge': ['1b', '2d'], 'Bi': ['2c'], 'Te': ['1a', '2d', '2d']} | 164 | −0.412 | 4 | 0.85 |
| 41 | Ge2Sb2Te8 | {'Ge': ['2c'], 'Sb': ['2d'], 'Te': ['1a', '2d', '2d', '3e']} | 164 | −0.387 | 4 | 0.74 |
| 42 | Ge3Bi2Te4 | {'Ge': ['1a', '2d'], 'Bi': ['2c'], 'Te': ['2d', '2d']} | 164 | −0.358 | 4 | 0.92 |
| 43 | Ge1Bi4Te5 | {'Ge': ['1a'], 'Bi': ['2d', '2d'], 'Te': ['1b', '2c', '2d']} | 164 | −0.357 | 4 | 0.87 |
| 44 | Ge2Bi4Te6 | {'Ge': ['2d'], 'Bi': ['2c', '2d'], 'Te': ['1b', '2d', '3e']} | 164 | −0.349 | 4 | 0.68 |
| 45 | Ge1Sb4Te7 | {'Ge': ['1b'], 'Sb': ['2d', '2d'], 'Te': ['2c', '2d', '3f']} | 164 | −0.345 | 4 | 0.60 |
| 46 | Ag2Bi3Te7 | {'Ag': ['2d'], 'Bi': ['1b', '2d'], 'Te': ['2d', '2d', '3e']} | 164 | −0.335 | 4 | 0.61 |
| 47 | Ge3Sb2Te5 | {'Ge': ['1b', '2c'], 'Sb': ['2d'], 'Te': ['1a', '2d', '2d']} | 164 | −0.318 | 4 | 0.60 |
| 48 | Ge2Bi5Te5 | {'Ge': ['2d'], 'Bi': ['2c', '3e'], 'Te': ['1b', '2d', '2d']} | 164 | −0.315 | 4 | 0.68 |
| 49 | Ge1Bi4Te4 | {'Ge': ['1b'], 'Bi': ['2d', '2d'], 'Te': ['2d', '2d']} | 164 | −0.305 | 4 | 0.96 |
| 50 | Ag2Bi3Te4 | {'Ag': ['2d'], 'Bi': ['1b', '2d'], 'Te': ['2d', '2d']} | 164 | −0.299 | 4 | 0.87 |
| 51 | Ag2Bi4Te4 | {'Ag': ['2d'], 'Bi': ['1a', '1b', '2d'], 'Te': ['2d', '2d']} | 164 | −0.278 | 4 | 0.85 |
| 52 | Ti3Sb3Te8 | {'Ti': ['1a', '2c'], 'Sb': ['1b', '2d'], 'Te': ['2c', '6i']} | 164 | −0.790 | 5 | 0.55 |
| 53 | Zr1Bi4Te6 | {'Zr': ['1a'], 'Bi': ['2d', '2d'], 'Te': ['2c', '2d', '2d']} | 164 | −0.677 | 5 | 0.77 |
| 54 | Ga4Bi2Te6 | {'Ga': ['1a', '1b', '2d'], 'Bi': ['2c'], 'Te': ['2c', '2d', '2d']} | 164 | −0.559 | 5 | 0.63 |
| 55 | Ga2Bi2Te6 | {'Ga': ['2d'], 'Bi': ['2c'], 'Te': ['2d', '2d', '2d']} | 164 | −0.506 | 5 | 0.77 |
| 56 | Sn1Pb4Te7 | {'Sn': ['1a'], 'Pb': ['2d', '2d'], 'Te': ['1b', '2c', '2d', '2d']} | 164 | −0.503 | 5 | 0.58 |
| 57 | Ge1Bi4Te7 | {'Ge': ['1a'], 'Bi': ['2c', '2d'], 'Te': ['1b', '2c', '2c', '2c']} | 164 | −0.485 | 5 | 0.66 |
| 58 | Ge2Bi2Te6 | {'Ge': ['2d'], 'Bi': ['2c'], 'Te': ['2c', '2d', '2d']} | 164 | −0.467 | 5 | 0.83 |
| 59 | Sn1Sb4Te7 | {'Sn': ['1b'], 'Sb': ['2d', '2d'], 'Te': ['1a', '2c', '2d', '2d']} | 164 | −0.410 | 5 | 0.52 |
| 60 | Ge2Sb2Te7 | {'Ge': ['2c'], 'Sb': ['2d'], 'Te': ['1a', '2c', '2d', '2d']} | 164 | −0.398 | 5 | 0.85 |
| 61 | Ge4Bi2Te5 | {'Ge': ['2c', '2d'], 'Bi': ['2c'], 'Te': ['1a', '2d', '2d']} | 164 | −0.393 | 5 | 0.96 |
| 62 | Ag3Bi3Te6 | {'Ag': ['1b', '2d'], 'Bi': ['1a', '2c'], 'Te': ['2c', '2c', '2d']} | 164 | −0.342 | 5 | 0.52 |
| 63 | Ag2Bi2Te10 | {'Ag': ['2d'], 'Bi': ['1a', '1b'], 'Te': ['2c', '2d', '6i']} | 164 | −0.338 | 5 | 0.55 |
| 64 | Ag4Bi2Te7 | {'Ag': ['2c', '2d'], 'Bi': ['2d'], 'Te': ['2c', '2d', '3e']} | 164 | −0.329 | 5 | 0.81 |
| 65 | Ge4Bi3Te4 | {'Ge': ['2c', '2d'], 'Bi': ['1b', '2c'], 'Te': ['2d', '2d']} | 164 | −0.294 | 5 | 0.93 |
| 66 | Rb4Sb4Se8 | {'Rb': ['4b'], 'Sb': ['4a'], 'Se': ['8e']} | 141 | −0.927 | 1 | 0.75 |
| 67 | Cu8Ge4Se16 | {'Cu': ['8d'], 'Ge': ['4b'], 'Se': ['16h']} | 141 | −0.534 | 2 | 0.83 |
| 68 | Ag4Bi8Se8 | {'Ag': ['4a'], 'Bi': ['8e'], 'Se': ['8e']} | 141 | −0.419 | 2 | 0.62 |
| 69 | Tb2Ag2Se4 | {'Tb': ['2c'], 'Ag': ['2b'], 'Se': ['4d']} | 129 | −1.467 | 1 | 0.56 |
| 70 | Cu2B6Se4 | {'Cu': ['2a'], 'B': ['2c', '4d'], 'Se': ['2b', '2c']} | 129 | −0.433 | 2 | 0.94 |
| 71 | Cu2Ag4Se4 | {'Cu': ['2a'], 'Ag': ['4f'], 'Se': ['2b', '2c']} | 129 | −0.271 | 2 | 0.89 |
| 72 | Cu4Ag4Se4 | {'Cu': ['2a', '2c'], 'Ag': ['4d'], 'Se': ['2b', '2c']} | 129 | −0.249 | 2 | 0.88 |
| 73 | Ag1Sb1Te2 | {'Ag': ['1c'], 'Sb': ['1b'], 'Te': ['1a', '1d']} | 123 | −0.207 | 0 | 0.84 |
| 74 | Ag3Sb1Te6 | {'Ag': ['1c', '2e'], 'Sb': ['1d'], 'Te': ['1a', '1b', '4i']} | 123 | −0.191 | 1 | 0.57 |
| 75 | Zn4Ga2Se4 | {'Zn': ['4d'], 'Ga': ['2b'], 'Se': ['4e']} | 121 | −0.823 | 1 | 0.87 |
| 76 | Zn4Ge2Se8 | {'Zn': ['4d'], 'Ge': ['2a'], 'Se': ['8i']} | 121 | −0.776 | 2 | 0.50 |
| 77 | Cu4In2Se8 | {'Cu': ['4d'], 'In': ['2b'], 'Se': ['8i']} | 121 | −0.511 | 2 | 0.82 |
| 78 | Cu6Sn2Se8 | {'Cu': ['2a', '4d'], 'Sn': ['2b'], 'Se': ['8i']} | 121 | −0.430 | 2 | 0.62 |
| 79 | Cu4Ge4Se8 | {'Cu': ['4d'], 'Ge': ['2a', '2b'], 'Se': ['8i']} | 121 | −0.426 | 2 | 0.60 |
| 80 | Cu6Sb2Se8 | {'Cu': ['2b', '4d'], 'Sb': ['2a'], 'Se': ['8i']} | 121 | −0.394 | 2 | 0.92 |
| 81 | Cu8Sn4Se8 | {'Cu': ['4c', '4d'], 'Sn': ['2a', '2b'], 'Se': ['8i']} | 121 | −0.386 | 2 | 0.68 |
| 82 | Cu4Ge6Se8 | {'Cu': ['4d'], 'Ge': ['2b', '4c'], 'Se': ['8i']} | 121 | −0.357 | 2 | 0.51 |
| 83 | Zn8Ge4Se8 | {'Zn': ['8f'], 'Ge': ['2a', '2b'], 'Se': ['8i']} | 121 | −0.704 | 3 | 0.92 |
| 84 | Cu6Sb2Se12 | {'Cu': ['2b', '4d'], 'Sb': ['2a'], 'Se': ['4e', '8i']} | 121 | −0.445 | 3 | 0.68 |
| 85 | Zn6Si8Se16 | {'Zn': ['2b', '4d'], 'Si': ['8f'], 'Se': ['8f', '8i']} | 121 | −0.811 | 4 | 0.73 |
| 86 | Zn4Ge10Se8 | {'Zn': ['4d'], 'Ge': ['2b', '8i'], 'Se': ['8i']} | 121 | −0.574 | 4 | 0.85 |
| 87 | Cu12Ge2Se16 | {'Cu': ['4d', '8f'], 'Ge': ['2b'], 'Se': ['8g', '8i']} | 121 | −0.544 | 4 | 0.89 |
| 88 | Cu4Sn2Se12 | {'Cu': ['4e'], 'Sn': ['2b'], 'Se': ['4e', '8i']} | 121 | −0.542 | 4 | 0.76 |
| 89 | Cu4Sn4Se16 | {'Cu': ['4d'], 'Sn': ['2a', '2b'], 'Se': ['8i', '8i']} | 121 | −0.539 | 4 | 0.67 |
| 90 | Cu12Sn2Se16 | {'Cu': ['4d', '8f'], 'Sn': ['2b'], 'Se': ['8f', '8i']} | 121 | −0.516 | 4 | 0.66 |
| 91 | Cu12Ge4Se8 | {'Cu': ['4d', '8i'], 'Ge': ['2a', '2b'], 'Se': ['8i']} | 121 | −0.358 | 4 | 0.64 |
| 92 | Zn16Ge4Se12 | {'Zn': ['16j'], 'Ge': ['2a', '2b'], 'Se': ['4d', '8i']} | 121 | −0.655 | 5 | 0.57 |
| 93 | Cu4Ge10Se20 | {'Cu': ['4d'], 'Ge': ['2b', '8h'], 'Se': ['4c', '8i', '8i']} | 121 | −0.583 | 5 | 0.56 |
| 94 | Cu12Sn6Se12 | {'Cu': ['4c', '4d', '4e'], 'Sn': ['2b', '4e'], 'Se': ['4e', '8i']} | 121 | −0.381 | 5 | 0.51 |
| 95 | Zn10Ga4Se10 | {'Zn': ['2a', '4e', '4e'], 'Ga': ['4f'], 'Se': ['2d', '4e', '4f']} | 119 | −0.646 | 5 | 0.63 |
| 96 | Na4Ge6Se6 | {'Na': ['4b'], 'Ge': ['2a', '4b'], 'Se': ['2a', '4b']} | 107 | −0.683 | 5 | 0.98 |
| 97 | Ge8Bi4Te12 | {'Ge': ['4a', '4c'], 'Bi': ['4c'], 'Te': ['4c', '4c', '4c']} | 63 | −0.501 | 5 | 0.61 |
| 98 | Cu2Sn2Se4 | {'Cu': ['2a'], 'Sn': ['2b'], 'Se': ['4d']} | 44 | −0.475 | 4 | 0.72 |
| 99 | Cu2Ge4Se4 | {'Cu': ['2a'], 'Ge': ['2b', '2b'], 'Se': ['4d']} | 44 | −0.376 | 5 | 0.72 |
| 10 | Cu1Ag1Te2 | {'Cu': ['1c'], 'Ag': ['1b'], 'Te': ['1a', '1a']} | 25 | −0.201 | 4 | 0.60 |
| 10 | Cu2Ag1Te2 | {'Cu': ['1a', '1c'], 'Ag': ['1a'], 'Te': ['1a', '1d']} | 25 | −0.200 | 5 | 0.73 |

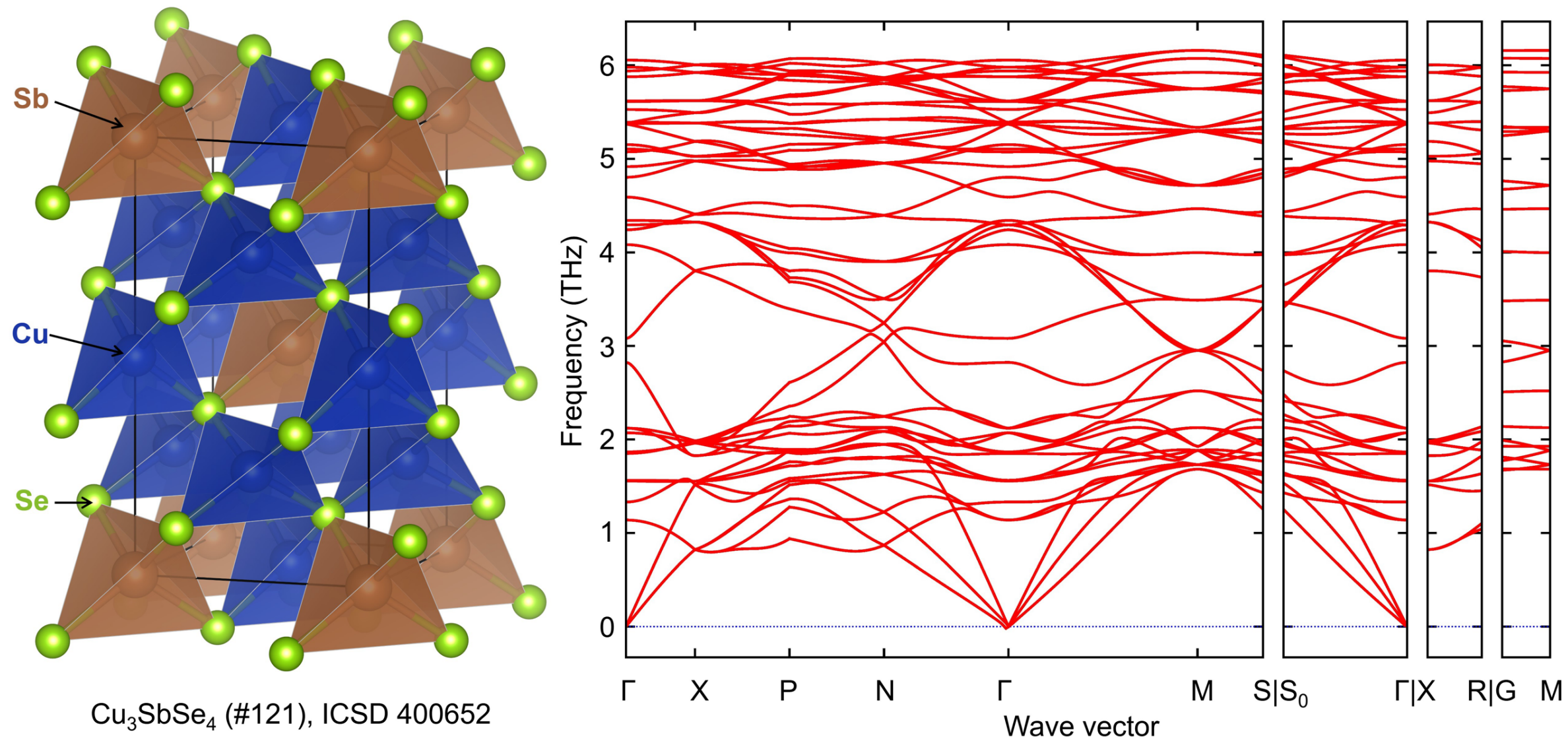

Figure S6. The structure (left) and phonon dispersion (right) of stannite $Cu_3SbSe_4$.

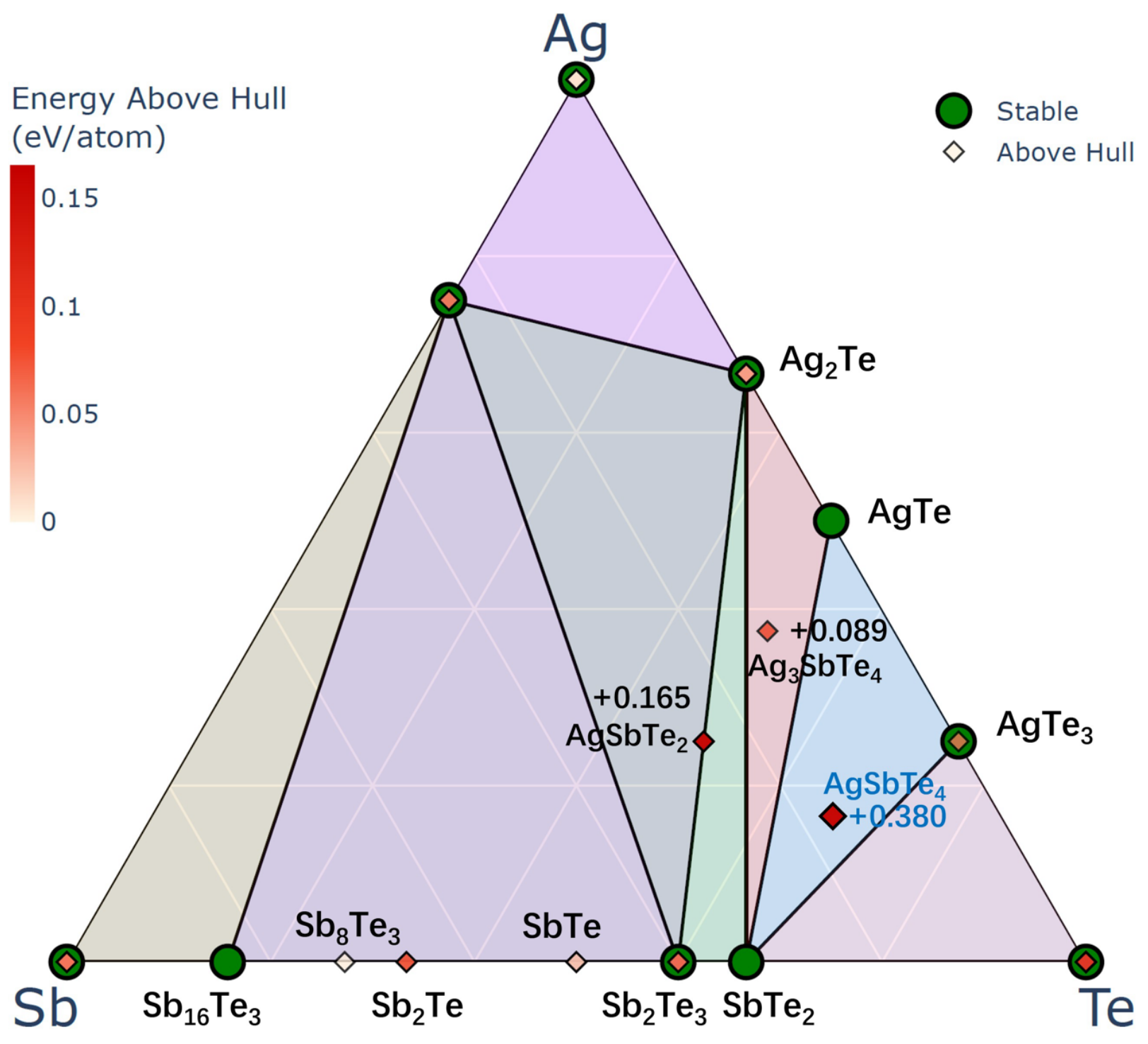

Figure S7. The Ag-Sb-Te phase diagram.

**10. Validity/Stability Metric in recent generative models.**

| Metrics used | Stability metric defined via Energy above the Convex Hull ($E_{hull}$) | Coordination Environments | Effort for Structure Relaxation | Wyckoff Site Symmetry within the unit cell for each Space Group | Phonon Stability |
|---|---|---|---|---|---|
| CrysTens | 0.115meV/atom (1000 samples test set) | / | / | Space group | / |
| CDVAE | / | Min (Interatomic distance) > 0.5 Å | / | / | / |
| MatterGen | < 0.1eV/atom | Min (Interatomic distance) > 0.5 Å | RMSD < 0.076 Å for 95 % of generated structures | Fine-tuning model on space Group symmetry only: ~20% success rate for space group and stability | / |
| DiffCSP | / | Min (Interatomic distance) > 0.5 Å | Match Rate, MR=52.02 | Space group symmetry only | / |
| DiffCSP++ | / | Min (Interatomic distance) > 0.5 Å | MR=52.17 | Space group and Wyckoff Site symmetry | / |
| WyCryst (Our Work) | Crystal Structure Prediction (CSP):<br>a. Ground State (lowest energy structure) and<br>b. Polymorph (particular space group) for any composition | Interatomic distance > (1.2-1.5)*atomic radius (by design)# | / | Space group and Wyckoff symmetry | ~8% on initial tests |

# This is through trial-and-error to capture most bonds appropriately.